\documentclass{amsart}

\usepackage{amsmath, amsthm, amssymb, amsbsy,amsfonts,amscd}
\usepackage{graphicx}
\usepackage{color}

\title[%
Extended Schur's $Q$-functions and the full Kostant--Toda hierarchy for type $D$
]{%
Extended Schur's $Q$-functions and the full Kostant--Toda hierarchy 
on the Lie algebra of type $D$
}

\author{%
Yuji Kodama and Soichi Okada
}

\date{\today}
\thanks{The second author was partially supported by JSPS KAKENHI 21K03202.}
\address{Department of Mathematics, Ohio State University,
Columbus, OH 43210, USA}
\email{kodama@math.ohio-state.edu}
\address{Graduate School of Mathematics, Nagoya University, Nagoya 464-8602, Japan}
\email{okada@math.nagoya-u.ac.jp}
\subjclass[2000]{}

\theoremstyle{definition}
\newtheorem{theorem}{Theorem}[section]

\newtheorem{definition}[theorem]{Definition}
\newtheorem{proposition}[theorem]{Proposition}
\newtheorem{lemma}[theorem]{Lemma}

\newtheorem{remark}[theorem]{Remark}

\newcommand\g{\mathfrak{g}}
\newcommand\n{\mathfrak{n}}
\newcommand\h{\mathfrak{h}}
\newcommand\bb{\mathfrak{b}}

\newcommand\GG{\mathcal{G}}
\newcommand\NN{\mathcal{N}}
\newcommand\HH{\mathcal{H}}
\newcommand\BB{\mathcal{B}}

\newcommand\Sym{\mathfrak{S}} 
\newcommand\WD{\mathfrak{D}}  
\newcommand\tv{\mathbf{t}}    
\DeclareMathOperator{\Lie}{Lie}
\DeclareMathOperator{\Ad}{Ad}
\DeclareMathOperator{\ad}{ad}
\DeclareMathOperator{\height}{ht}
\DeclareMathOperator{\wt}{wt}
\DeclareMathOperator{\diag}{diag}
\DeclareMathOperator{\Pf}{Pf}
\DeclareMathOperator{\tr}{tr}

\newcommand\GL{\mathbf{GL}}
\newcommand\gl{\mathfrak{gl}}
\newcommand\SL{\mathbf{SL}}
\newcommand\spl{\mathfrak{sl}}
\newcommand\Orth{\mathbf{O}}
\newcommand\SOrth{\mathbf{SO}}
\newcommand\sorth{\mathfrak{so}}
\newcommand\Spin{\mathbf{Spin}}

\newcommand\Comp{\mathbb{C}}

\newcommand\Int{\mathbb{Z}}

\newcommand\ep{\varepsilon}
\renewcommand\tilde{\widetilde}
\renewcommand\hat{\widehat}
\newcommand\trans{{}^t\!}

\newcommand\twedge{{\textstyle\bigwedge}} 

\numberwithin{equation}{section}

\begin{document}

\dedicatory{\hfill Dedicated to the memory of Hermann Flaschka}

\begin{abstract}
The full Kostant--Toda hierarchy on a semisimple Lie algebra is 
a system of Lax equations, in which the flows are determined by 
the gradients of the Chevalley invariants.
This paper is concerned with the full Kostant--Toda hierarchy on 
the even orthogonal Lie algebra.
By using a Pfaffian of the Lax matrix as one of the Chevalley invariants, 
we construct an explicit form of the flow associated to this invariant.
As a main result, we introduce an extension of the Schur's $Q$-functions 
in the time variables, and use them to give explicit formulas 
for the polynomial $\tau$-functions of the hierarchy. 
\end{abstract}

\maketitle

\setcounter{tocdepth}{1}
\tableofcontents

\section{%
Introduction
}
\label{sec:introduction}

We start with fixing some notations.
Let $\g$ be a finite-dimensional complex semisimple Lie algebra of rank $n$.
We fix a triangular decomposition
\[
\g = \overline{\n} \oplus \h \oplus \n,
\]
where $\h$ is a Cartan subalgebra and $\n$ (resp. $\overline{\n}$) 
is the nilradical of a Borel subalgebra $\bb = \h \oplus \n$ 
(resp. $\overline{\bb} = \h \oplus \overline{\n}$).
Let $\Sigma \subset \h^*$ be the root system of $\g$ with respect to $\h$, 
and $\Sigma^+$ and $\Pi = \{ \alpha_1, \dots, \alpha_n \}$ 
the positive system and the simple system associated with $\n$ respectively.
Let $\kappa : \g \times \g \to \Comp$ be an invariant non-degenerate symmetric bilinear form 
on $\g$.
If $\g$ is simple, then $\kappa$ is proportional to the Killing form of $\g$.
Let $\{ H_i : 1 \le i \le n \} \cup \{ X_\alpha : \alpha \in \Sigma \}$ 
be a Chevalley basis satisfying
\begin{gather*}
[H_i, H_j] = 0 \quad(1 \le i, \, j \le n),
\\
[H_i, X_\alpha] = \alpha(H_i) X_\alpha \quad(1 \le i \le n, \, \alpha \in \Sigma),
\\
[X_\alpha, X_{-\alpha}] = H_\alpha \quad(\alpha \in \Sigma),
\\
[X_\alpha, X_\beta] = N_{\alpha,\beta} X_{\alpha+\beta} \quad(\alpha, \beta \in \Sigma),
\end{gather*}
where $H_\alpha \in \h$ 
and we use the convention $X_{\alpha+\beta} = 0$ if $\alpha+\beta \not\in \Sigma$.
Let $\GG$ be a connected complex semisimple Lie group 
with Lie algebra $\Lie \GG = \g$.

The \emph{full Kostant--Toda hierarchy} (f-KT hierarchy for short) 
on the Lie algebra $\g$ is defined as follows.
Let $\tv = (t_1, t_2, \dots, t_n)$ be time variables and 
consider an element $L(\tv) \in \g$ of the form
\begin{equation}
\label{eq:Lax}
L(\tv)
 = 
\sum_{i=1}^n a_i(\tv) H_i
 + 
\sum_{i=1}^n X_{\alpha_i}
 +
\sum_{\alpha \in \Sigma^+} b_{\alpha}(\tv) X_{-\alpha},
\end{equation}
which is called the \emph{Lax matrix}.
Here  $a_i(\tv)$ and $b_{\alpha}(\tv)$ are functions of the multi-time variables $\tv$.
It is known as Chevalley's theorem that 
the ring $\Comp[\g]^{\GG}$ of $\GG$-invariant polynomial functions on $\g$ 
is generated by $n$ algebraically independent homogeneous polynomials 
$I_1, \dots, I_n$, 
which are referred to as the fundamental invariants or the Chevalley invariants.
Then the f-KT hierarchy is defined by
\begin{equation}
\label{eq:fKT}
\frac{\partial L}{\partial t_k} 
 =
\left[ P_k^{\ge 0}, L \right]
\quad\text{with}\quad
P_k = (\nabla I_k)(L)
\qquad
(1 \le k \le n),
\end{equation}
where $\nabla f : \g \to \g$ denotes the gradient of $f \in \Comp[\g]$ given by
\[
\kappa \left( \nabla f(X), Z \right)
 =
\frac{d}{dt} f(X + t Z) \Big|_{t=0}
\qquad(Z \in \g),
\]
and $P_k^{\ge 0}$ is the image of $P_k$ under the projection onto $\bb$ with respect to 
the direct sum decomposition $\g = \bb \oplus \overline{\n}$.

The full Kostant--Toda lattice, the first member of the hierarchy 
with $I_1 = \frac{1}{2} \tr(L^2)$ (i.e. $P_1 = L$) on $\g = \spl_{n+1}(\Comp)$, 
was first introduced by Ercolani, Flaschka and Singer in \cite{EFS:93}, 
where the main purpose of their paper is to show its complete integrability 
by constructing a sufficient number of the first integrals, 
called \emph{chop} integrals. 
The integrability of the lattice for other types of semisimple Lie algebras 
was shown by Gekhtman and Shapiro \cite{GS:99}, 
in which they gave a Lie theoretic meaning of the chop integrals 
and showed that the generic orbits of the f-KT lattices are 
completely integrable in a noncommutative sense.

In \cite{Kos:79, GW:84}, the f-KT lattice with tridiagonal (or Jacobi) element $L$ 
of a semi-simple Lie algebra $\g$, called simply the Kostant--Toda (KT) lattice, 
is studied by virtue of the representation theory. 
The main result of Kostant \cite{Kos:79} is to show that the KT lattice 
is a completely integrable Hamiltonian system 
and the integration of the KT lattice is completely determined 
by the weight structure of the fundamental representations of the corresponding group $\GG$. 
More precisely, the integration turns out to be an Iwasawa-type factorization problem.
Then Goodman and Wallach \cite{GW:84} found formulas of the solutions 
in terms of the $\tau$-functions which are given by
\[
\tau_i(t)
 =
\langle v_{\varpi_i},\, g(t) \cdot v_{\varpi_i} \rangle
\qquad(1 \le i \le n),
\]
where $\langle \cdot,\cdot \rangle$ is a non-degenerate bilinear form 
on the irreducible highest weight representation corresponding to 
the fundamental weight $\varpi_i$ with highest weight vector $v_{\varpi_i}$.
Here the group element $g(t) \in \GG$ is defined by 
$g(t) = \exp (tL(0))$ with the initial matrix $L(0)$. 
In this paper, we extend this formula to the $\tau$-function for the f-KT hierarchy 
on $\g = \sorth_{2n}(\Comp)$, and give all the polynomial $\tau$-functions.

There have been considerable interest on the singular solutions 
of the f-KT hierarchy. 
The singular structures are determined by the zeros of the $\tau$-functions, 
called the Painlev\'e divisors in \cite{FH:91} (see also \cite{CK:06}). 
Around a point on the Painlev\'e divisor, the $\tau$-function admits 
a power series expansion in $\tv = (t_1, \dots, t_n)$, 
whose leading order term is given by a Schur-type polynomial. 
In \cite{FH:91}, Flaschka and Heine determine the obstructions of 
the Gauss decomposition and link the result to the singularity of the KT lattice.

The Schur-type functions also appear in the $\tau$-functions 
of the KP-type hierarchies \cite{You:89, Shi:16, KL:19, HO:21,Le:22}.
In particular, You \cite{You:89} gives the polynomial $\tau$-functions 
of the BKP hierarchy in terms of Schur's $Q$ functions, 
which were first introduced by Schur in the study of projective representations 
of symmetric group. 
In this paper, we will introduce an extension of Schur's $Q$-functions, 
and show that these extensions give the polynomial $\tau$-functions 
to the f-KT hierarchy on $\g = \sorth_{2n}(\Comp)$. 

You \cite{You:89, You:92} also gives identities relating $2$-reduced Schur functions 
and Schur's $Q$-functions (see also \cite{JP:91}). 
In general, any $2$-reduced Schur function can be expressed as a sum of products of pais of 
Schur's $Q$-functions (see \cite{P:98, HO:21a}).
In the course of the proof of our main thoerem, we see that some of these identities 
are interpreted as relations among matrix coefficients involving the half-spin representations 
of the spin group $\Spin_{2n}(\Comp)$.

Our main goal in this paper is to give a complete list of the rational solutions 
(i.e. the polynomial $\tau$-functions) to the f-KT hierarchy on the Lie algebra 
$\g = \sorth_{2n}(\Comp)$. 
Sections~\ref{sec:fKT} and \ref{sec:tau} discuss general features of the f-KT hierarchy.
In Section~\ref{sec:fKT}, we prove that the flows in \eqref{eq:fKT} mutually commute 
and that any $\mathcal{G}$-invariant polynomial is a first integral of the f-KT hierarchy 
(Proposition~\ref{prop:fKT}). 
In Section~\ref{sec:tau}, we show that the solutions of the f-KT hierarchy 
is completely determined and expressed by the $\tau$-functions, 
which are defined as the matrix coefficients on the irreducible highest weight representations 
of $\GG$ (Definition~\ref{def:tau} and Proposition~\ref{prop:solution}). 
We also give the formula of polynomial $\tau$-functions (Proposition~\ref{prop:nilpotent}).
Sections~\ref{sec:fKT_D} and \ref{sec:Psolutions} focus on the f-KT hierarchy 
on the Lie algebra $\sorth_{2n}(\Comp)$.
In Section~\ref{sec:fKT_D}, 
by choosing the specific Chevalley invariants including a Pfaffian invariant,
we give an explicit formula of the f-KT hierarchy on $\sorth_{2n}(\Comp)$ 
(Proposition~\ref{prop:fKT_D}).
In particular, one of the Chevalley invariants is given by a Pfaffian of the Lax matrix, 
and the flow associated with the Pfaffian invariant has not been found explicitly 
in the previous works. 
In Section~\ref{sec:Psolutions}, we introduce an extension of Schur's $Q$-functions 
in the time variables $\tv = (t_1, t_3, \dots, t_{2n-3}, s)$, 
where the variable $s$ is the flow parameter corresponding to the Chevalley invariant 
given by Pfaffian. 
Our main result (Theorem~\ref{thm:tau}) provides explicit formulas 
for the polynomial $\tau$-functions in terms of extended Schur's $Q$-functions.
\section{%
The full Kostant--Toda hierarchy
}
\label{sec:fKT}

In this section we prove that 
the flows generated by the Chevalley invariants in the f-KT hierarchy \eqref{eq:fKT} 
mutually commute 
and that any $\GG$-invariant polynomial is a first integral of the f-KT hierarchy.

Let $\{ Z_1, \dots, Z_n \}$ be a basis of $\g$ and $\{ \zeta_1, \dots, \zeta_n \}$ 
be the dual basis of $\g^*$.
We can identify a polynomial function $f \in \Comp[\g]$ with 
a polynomial in $\zeta_1, \dots, \zeta_N$.
Then we have
\begin{equation}
\label{eq:grad}
(\nabla f) (X) = \sum_{i=1}^N \frac{\partial f}{\partial \zeta_i}(X) Z^i,
\end{equation}
where $\{ Z^1, \dots, Z^N \}$ is the dual basis of $\g$ with respect to 
the bilinear form $\kappa$.

In what follows, we choose the quadratic Casimir element as the first member $I_1$ 
of the Chevalley invariants, 
i.e., we put
\[
I_1 = \frac{1}{2} \sum_{i=1}^N \zeta_i \zeta^i,
\]
where $\{ \zeta^1, \dots, \zeta^N \}$ is the basis of $\g^*$ 
dual to $\{ Z^1, \dots, Z^N \}$.
Since $(\nabla I_1) (X) = X$, the first member of the hierarchy \eqref{eq:fKT}, 
called the f-KT lattice, is 
given by
\begin{equation}
\label{eq:fKT1}
\frac{\partial L}{\partial t_1} = \left[ L^{\ge 0}, L \right].
\end{equation}

Now we have the following proposition.
\begin{proposition}
\label{prop:fKT}
Let $L$ be a solution to the f-KT hierarchy \eqref{eq:fKT}.
\begin{enumerate}
\item[(1)]
We have
\[
\frac{\partial^2 L}
     {\partial t_k \partial t_l}
 =
\frac{\partial^2 L}
     {\partial t_l \partial t_k}
\qquad (1 \le k, \, l \le n).
\]
Hence the fKT flows commute with each other.
\item[(2)]
If $f \in \Comp[\g]^\GG$, then we have
\[
\frac{ \partial }{ \partial t_k } f(L) = 0
\qquad (1 \le k \le n).
\]
Hence $f(L)$ is invariant under the f-KT flows.
\end{enumerate}
\end{proposition}

In the proof of this proposition, we use the following properties of gradients.

\begin{lemma}
\label{lem:grad}
\begin{enumerate}
\item[(1)]
For $f \in \Comp[\g]^\GG$, $g \in \GG$ and $X \in \g$, we have
\begin{equation}
\label{eq:grad1}
\Ad(g) (\nabla f)(X) = (\nabla f) (\Ad(g) X).
\end{equation}
\item[(2)]
For $f \in \Comp[\g]^\GG$ and $X$, $Y \in \g$, we have
\begin{equation}
\label{eq:grad2}
\left[ (\nabla f) (X), Y \right]
 =
\sum_{i,j=1}^N
 \frac{ \partial^2 f }{ \partial \zeta_i \partial \zeta_j }(X)
 \zeta_j([X,Y]) Z^i.
\end{equation}
\item[(3)]
For $f \in \Comp[\g]^\GG$ and $X \in \g$, we have
\begin{equation}
\label{eq:grad3}
\left[ (\nabla f) (X), X \right] =0.
\end{equation}
\item[(4)]
For $f$, $g \in \Comp[\g]^\GG$ and $X \in \g$, we have
\begin{equation}
\label{eq:grad4}
\left[ (\nabla f) (X), (\nabla g) (X) \right] = 0.
\end{equation}
\end{enumerate}
\end{lemma}

\begin{proof}
(1)
Since $\kappa : \g \times \g \to \Comp$ is an invariant bilinear form 
and $f$ is $\Ad(\GG)$-invariant, we have
\begin{align*}
\kappa \left( \Ad(g^{-1}) (\nabla f) ( \Ad(g) X), Z \right)
 &=
\kappa \left( (\nabla f) ( \Ad(g) X), \Ad(g) Z \right)
 =
\frac{d}{dt}
f ( \Ad(g) (X + t Z) )
\Big|_{t=0}
\\
 &=
\frac{d}{dt}
f ( X + t Z )
\Big|_{t=0}
 =
\kappa \left( (\nabla f) (X), Z \right)
\end{align*}
for any $Z \in \g$.
Hence we have $\Ad(g^{-1}) (\nabla f) ( \Ad(g) X) = (\nabla f)(X)$, i.e. 
$(\nabla f) (\Ad(g) X) = \Ad(g) (\nabla f)(X)$.

(2)
We apply (1) to $g = \exp (t \ad Y) \in \GG$.
Then by using \eqref{eq:grad} we have
$$
\sum_{i=1}^N \frac{\partial f}{\partial \zeta_i} \left( \exp (t \ad Y) X \right) Z^i
 =
\exp (t \ad Y) \left( \nabla f(X) \right).
$$
By differentiating the both sides with respect to $t$ and putting $t=0$, 
we obtain
$$
\sum_{i,j=1}^N
 \frac{ \partial^2 f }{ \partial \zeta_i \partial \zeta_j }(X)
 \zeta_j([Y,X]) Z^i
 =
[Y, (\nabla f)(X)].
$$

(3) is obtained from (2) by putting $Y = X$.

(4) follows from (2) by specializing $Y = (\nabla g)(X)$ and using (3).
\end{proof}

We can use this lemma to prove Proposition~\ref{prop:fKT}.

\begin{proof}[Proof of Proposition~\ref{prop:fKT}]
(1)
Recall $P_k = (\nabla I_k)(L)$ and $P_k^{\ge 0}$ is the image of $P_k$ 
under the projection from $\g$ to $\bb$.
By using \eqref{eq:fKT} and the Jacobi identity, we have
\begin{align*}
\frac{\partial^2 L}{\partial t_l \partial t_k}
 -
\frac{\partial^2 L}{\partial t_k \partial t_l}
 &=
\left[ \frac{\partial P_k^{\ge 0}}{\partial t_l}, L \right]
 +
\left[ P_k^{\ge 0}, \left[ P_l^{\ge 0},L \right] \right]
 - 
\left[ \frac{\partial P_l^{\ge 0}}{\partial t_k}, L \right]
 -
\left[ P_l^{\ge 0}, \left[ P_k^{\ge 0},L \right] \right]
\\
 &=
\left[
 \frac{\partial P_k^{\ge 0}}{\partial t_l}
 - \frac{\partial P_l^{\ge 0}}{\partial t_k}
 + \left[ P_k^{\ge 0}, P_l^{\ge 0} \right], 
 L
\right].
\end{align*}
So it is enough to prove
\begin{equation}
\label{eq:ZS}
\frac{ \partial P_k^{\ge 0}}{ \partial t_l }
 -
\frac{ \partial P_l^{\ge 0}}{ \partial t_k }
 +
\left[ P_k^{\ge 0}, P_l^{\ge 0} \right]
 =
0.
\end{equation}

First we show that
\begin{equation}
\label{eq:ZS1}
\frac{ \partial P_k }{ \partial t_l } = \left[ P_l^{\ge 0}, P_k \right],
\quad
\frac{ \partial P_l }{ \partial t_k } = \left[ P_k^{\ge 0}, P_l \right].
\end{equation}
By using \eqref{eq:grad}, \eqref{eq:fKT} and \eqref{eq:grad2}, we obtain
\begin{align*}
\frac{ \partial P_k }{ \partial t_l }
 &=
\sum_{i,j=1}^N
 \frac{ \partial^2 I_k }{ \partial \zeta_i \partial \zeta_j }(L)
 \frac{ \partial \zeta_j(L)}{\partial t_l}
 Z^i
 =
\sum_{i,j=1}^N
 \frac{ \partial^2 I_k }{ \partial \zeta_i \partial \zeta_j }(L)
 \zeta_j([P_l^{\ge 0}, L])
 Z^i
\\
 &=
\left[
 P_l^{\ge 0}, (\nabla I_k)(L)
\right]
 =
\left[
 P_l^{\ge 0}, P_k
\right].
\end{align*}

We denote by $P_k^{<0}$ and $P_l^{<0}$ the images of $P_k$ and $P_l$ 
under the projection onto $\overline{\n}$ respectively.
Then $P_k = P_k^{\ge 0} + P_k^{<0}$, $P_l = P_l^{\ge 0} + P_l^{<0}$, 
and we can rewrite \eqref{eq:ZS1} as
\begin{gather*}
\frac{\partial P_k^{\ge 0}}{\partial t_l} + \frac{\partial P_k^{<0}}{\partial t_l}
 =
\left[ P_l^{\ge 0}, P_k^{\ge 0} \right] + \left[ P_l^{\ge 0}, P_k^{<0} \right],
\\
\frac{\partial P_l^{\ge 0}}{\partial t_k} + \frac{\partial P_l^{<0}}{\partial t_k}
 =
\left[ P_k^{\ge 0}, P_l^{\ge 0} \right] + \left[ P_k^{\ge 0}, P_l^{<0} \right].
\end{gather*}
Hence we have
\[
\frac{\partial P_k^{\ge 0}}{\partial t_l}
 - \frac{\partial P_l^{\ge 0}}{\partial t_k}
 + \left[ P_k^{\ge 0}, P_l^{\ge 0} \right]
 =
\frac{\partial P_l^{<0}}{\partial t_k}
 - \frac{\partial P_k^{<0}}{\partial t_l}
 - \left[ P_k^{\ge 0}, P_l^{\ge 0} \right]
 - \left[ P_k^{<0},P_l^{\ge 0} \right]
 - \left[ P_k^{\ge 0},P_l^{<0} \right].
\]
Since $[P_k,P_l] = 0$ by \eqref{eq:grad4}, we have
\[
\frac{\partial P_k^{\ge 0}}{\partial t_l}
 - \frac{\partial P_l^{\ge 0}}{\partial t_k}
 + \left[ P_k^{\ge 0}, P_l^{\ge 0} \right]
 =
\frac{\partial P_l^{<0}}{\partial t_k}
 - \frac{\partial P_k^{<0}}{\partial t_l}
 + \Big[ P_k^{<0}, P_l^{<0} \Big].
\]
Since the left hand side is an element of $\bb$ and 
the right hand side is an element of $\overline{\n}$, 
we obtain the desired identity \eqref{eq:ZS} 
and complete the proof of (1).

(2)
Since $\frac{\partial L}{\partial t_k} = \left[ P_k^{\ge 0}, L \right]$, 
we see
$$
\frac{ \partial f(L) }{ \partial t_k }
 =
\sum_{i=1}^N
 \frac{ \partial f }{ \partial \zeta_i}(L)
 \frac{ \partial \zeta_i(L) }{ \partial t_k }
 =
\sum_{i=1}^N
 \frac{ \partial f }{ \partial \zeta_i}(L)
 \zeta_i( [P_k^{\ge 0}, L] ).
$$
By using \eqref{eq:grad} and $[P_k^{\ge 0},L] = \sum_{i=1}^N \zeta_i([ P_k^{\ge 0},L]) Z_i$,
we have
\[
\frac{ \partial f(L) }{ \partial t_k }
 =
\kappa \left( (\nabla f)(L), [P_k^{\ge 0}, L] \right)
 =
- \kappa \left( [(\nabla f)(L),L], P_k^{\ge 0} \right).
\]
By \eqref{eq:grad3}, we conclude $\frac{ \partial f(L) }{ \partial t_k } = 0$.
\end{proof}

\begin{remark}
The Kostant--Toda (KT) lattice discussed in \cite{Kos:79} 
is a special case of \eqref{eq:fKT} 
with $b_{\alpha}(\tv) = 0$ for all $\alpha \in \Sigma^+ \setminus \Pi$, 
that is, the case where the Lax matrix $L$ is a Jacobi element of $\g$. 
The hierarchy of this type is also referred to as the tri-diagonal f-KT hierarchy, 
and the KT lattice (the first member of the hierarchy) is given by
\[
\frac{\partial a_i}{\partial t_1}
 =
b_{\alpha_i},
\quad
\frac{\partial b_{\alpha_i}}{\partial t_1}
 =
- \sum_{j=1}^n (C_{i,j} a_j) b_{\alpha_i}
\qquad (1 \le i \le n),
\]
where $C = \left( C_{i,j} \right) = \left( \alpha_i(H_j) \right)$ 
is the Cartan matrix of $\g$. 
The complete integrability of the lattice is shown 
by using the existence of the Chevalley invariants.
\end{remark}
\section{%
Solution method and the $\tau$-functions
}
\label{sec:tau}

In this section, we introduce the $\tau$-functions 
and prove that the solutions of the f-KT hierarchy can be described in terms of them.
We fix some notations. 
Recall $\GG$ is a connected complex semisimple Lie group with Lie algebra $\Lie(\GG) = \g$, 
and denote by $\HH$, $\NN$, $\overline{\NN}$, $\BB$ and $\overline{\BB}$ 
the connected subgroups of $\GG$ with Lie algebras $\h$, $\n$, 
$\overline{\n}$, $\bb$ and $\overline{\bb}$ respectively.
Let $W = N_{\GG}(\HH)/\HH$ be the Weyl group of $\g$, 
where $N_{\GG}(\HH)$ is the normalizer of $\HH$ in $\GG$, 
and fix a complete set of coset representatives $\{ \dot{w} : w \in W \}$.

\subsection{%
Matrix coefficients
}

Let $\theta : \GG \to \GG$ be the Chevalley involution satisfying
\begin{gather*}
\theta(h) = h^{-1} \quad(h \in \HH),
\\
\theta(\NN) = \overline{\NN},
\quad
\theta(\overline{\NN}) = \NN.
\end{gather*}
We denote by $(\rho(\lambda),V(\lambda))$ the irreducible highest weight representation 
of the group $\GG$ with highest weight $\lambda$ and highest weight vector $v_\lambda$. 
Then the contragredient representation $V(\lambda)^*$ is equivalent to 
the representation $(\rho_\lambda^\theta, V(\lambda)^\theta)$ of $\GG$ 
obtained from $V(\lambda)$ by twisting by $\theta$, 
i.e., $V(\lambda)^\theta = V(\lambda)$ and $\rho_\lambda^\theta = \rho_\lambda \circ \theta$.

\begin{lemma}
\label{lem:bilinear}
For each irreducible highest weight representation $V(\lambda)$ of $\GG$, 
there exists a unique bilinear form 
$\langle \cdot, \cdot \rangle_\lambda : V(\lambda) \times V(\lambda) \to \Comp$ 
satisfying
\begin{gather*}
\langle \theta(g) v, g w \rangle_\lambda = \langle v, w \rangle_\lambda
\quad(g \in G, \, v, w \in V),
\\
\langle v_\lambda, v_\lambda \rangle_\lambda = 1.
\end{gather*}
Moreover, 
there exists an orthonormal basis of $V(\lambda)$ consisting of weight vectors.
\end{lemma}

\begin{proof}
Since $V(\lambda)^* \cong V(\lambda)^\theta$, 
the canonical pairing $V(\lambda)^* \times V(\lambda) \to \Comp$ induces 
a desired bilinear form, and the uniqueness follows from Schur's Lemma.
The latter claim can be proved by showing that weight vectors of different weights 
are orthogonal with respect to $\langle \cdot, \cdot \rangle_\lambda$.
\end{proof}

\begin{definition}
\label{def:mat_coeff}
Given a dominant weight $\lambda$ of $\GG$, 
we define a function $c_\lambda : \GG \to \Comp$ by putting
\begin{equation}
\label{eq:def_mat_coeff}
c_\lambda(g)
 = 
\langle v_\lambda, g \cdot v_\lambda \rangle_\lambda
\qquad(g \in \GG),
\end{equation}
where $\langle \cdot, \cdot \rangle_\lambda$ is the bilinear form on $V(\lambda)$ 
given in Lemma~\ref{lem:bilinear}. That is, $c_\lambda(g)$ is a matrix coefficient for the irreducible highest weight representation of $\mathcal{G}$.
\end{definition}

The following lemma is useful.

\begin{lemma}
\label{lem:mat_coeff}
\begin{enumerate}
\item[(1)]
If $\{ u_1, \dots, u_r \}$ is an orthonormal basis of $V(\lambda)$, 
then we have for $v$, $w \in V(\lambda)$,
\begin{equation}
\label{eq:mat_coeff1}
\langle v, gh \cdot w \rangle_\lambda
 =
\sum_{i=1}^r
 \langle v, g \cdot u_i \rangle_\lambda
 \langle u_i, h \cdot w \rangle_\lambda.
\end{equation}
\item[(2)]
If $g \in \GG$ is written as $g = \overline{n} h n$ 
with $\overline{n} \in \overline{\NN}$, $h \in \HH$ and $n \in \NN$, 
then we have
\begin{equation}
\label{eq:mat_coeff2}
c_\lambda(\overline{n} h n) = \chi^\lambda(h),
\end{equation}
where $\chi^\lambda: \HH \to \Comp^\times$ is the character 
corresponding to $\lambda$.
\item[(3)]
For dominant weights $\lambda$ and $\mu$, we have
\begin{equation}
\label{eq:mat_coeff3}
c_\lambda(g) \cdot c_\mu(g) = c_{\lambda+\mu}(g)
\qquad(g \in \GG).
\end{equation}
\end{enumerate}
\end{lemma}

\begin{proof}
(1) 
Using the expansion $h \cdot w = \sum_{i=1}^r \langle u_i, h \cdot w \rangle_\lambda u_i$, 
we have
\[
\langle v, gh \cdot w \rangle_\lambda
 =
\langle v, g \cdot (h \cdot w) \rangle_\lambda
 =
\left\langle 
 v, 
 g\cdot \left(\sum_{i=1}^r \langle u_i, h \cdot w \rangle_\lambda u_i \right)
\right\rangle_\lambda
 =
\sum_{i=1}^r
 \langle u_i, h \cdot w \rangle_\lambda 
 \langle v, g \cdot u_i \rangle_\lambda,
\]
which is the desired formula.

(2)
Since $\theta(\overline{n})^{-1} \cdot v_\lambda = v_\lambda$, 
$n \cdot v_\lambda = v_\lambda$ and $h \cdot v_\lambda = \chi^\lambda(h) v_\lambda$, 
we have
\[
c_\lambda(\overline{n} h n)
 =
\langle v_\lambda, \overline{n} h n \cdot v_\lambda \rangle_\lambda
 =
\langle
 \theta(\overline{n})^{-1} \cdot v_\lambda, 
 h n \cdot v_\lambda
\rangle_\lambda
 =
\chi^\lambda(h) \langle v_\lambda, v_\lambda \rangle_\lambda
 =
\chi^\lambda(h).
\]

(3)
Since $V(\lambda+\mu)$ appears in the tensor product $V(\lambda) \otimes V(\mu)$ 
with multiplicity one, 
we obtain the embedding $\iota : V(\lambda+\mu) \to V(\lambda) \times V(\mu)$ 
such that $\iota(v_{\lambda+\mu}) = v_\lambda \otimes v_\mu$.
Then the restriction of the tensor product bilinear form 
$\langle \cdot, \cdot \rangle_\lambda \otimes \langle \cdot, \cdot \rangle_{\mu}$ 
to $V(\lambda+\mu)$ coincides with $\langle \cdot, \cdot \rangle_{\lambda+\mu}$ 
by the uniqueness in Lemma~\ref{lem:bilinear}.
Hence we have
\begin{align*}
c^{\lambda+\mu}(g)
 &=
\langle v^{\lambda+\mu}, g\cdot v^{\lambda+\mu} \rangle_{\lambda+\mu}
 =
\langle v^{\lambda} \otimes v^{\mu}, g \cdot (v^\lambda \otimes v^\mu) \rangle
 =
\langle v^\lambda \otimes v^\mu, (g \cdot v^\lambda) \otimes (g \cdot v^\mu) \rangle
\\
 &=
\langle v^\lambda, g \cdot v^\lambda \rangle_\lambda
\langle v^\mu, g \cdot v^\mu \rangle_\mu
 =
c^\lambda(g) c^\mu(g).
\end{align*}
\end{proof}

\subsection{%
$\tau$-functions and matrix factorization
}

To find a solution of the f-KT hierarchy \eqref{eq:fKT}, let us define (see also \cite{KW:15})
\begin{equation}
\label{eq:Theta}
\Theta(X;\tv)
 =
\sum_{k=1}^n t_k (\nabla I_k)(X)
\end{equation}
for $X \in \g$, 
and consider the exponential $\exp \Theta(X;\tv) \in \GG$.
Then the $\tau$-functions are introduced as follows.
Let $\varpi_1, \dots, \varpi_n$ be the fundamental weights of $\g$,
i.e. $\varpi_i(H_j) = \delta_{i,j}$ for $1 \le i, \, j \le n$.

\begin{definition}
\label{def:tau}
Suppose that $\GG$ is a connected, simply-connected complex semisimple Lie group 
with $\Lie \GG = \g$.
Then we define $\tau_k(X;\tv) = \tau_k(X;t_1, \dots, t_n)$ 
by the matrix coefficient $c_{\varpi_k}(g)$ with $g=\exp\Theta(X;\tv)$, i.e.
\begin{equation}
\label{eq:def_tau}
\tau_k(X;\tv)
 = 
c_{\varpi_k}( \exp \Theta(X;\tv) )
 =
\langle v_{\varpi_k}, \exp \Theta(X;t) \cdot v_{\varpi_k} \rangle_{\varpi_k}.
\end{equation}
More generally, for a vector $u \in V(\varpi_k)$, we define
\begin{equation}
\label{eq:def_gen_tau}
\tau_k(X;\tv;u)
 = 
\langle v_{\varpi_k}, \exp \Theta(X;t) \cdot u \rangle_{\varpi_k}.
\end{equation}
\end{definition}

We shall prove that the solutions of \eqref{eq:fKT} are described 
in terms of the $\tau$-functions.
First we show that the coefficients $b_\alpha$ of a solution 
to \eqref{eq:fKT} are uniquely determined by $a_1, \dots, a_n$.

\begin{proposition}
\label{prop:sol_b}
If $L = L(\tv)$ is a solution of the form \eqref{eq:Lax} 
to the full Kostant--Toda hierarchy \eqref{eq:fKT}, 
then $b_\alpha$'s ($\alpha \in \Sigma^+$) 
are expressed as polynomials in $a_i$'s and their derivatives.
In particular, we have
\begin{equation}
\label{eq:b=a'}
b_{\alpha_i} = \frac{\partial a_i}{\partial t_1}
\qquad(1 \le i \le n).
\end{equation}
\end{proposition}

\begin{proof}
Recall that the height $\height(\alpha)$ of a root 
$\alpha = \sum_{i=1}^n c_i \alpha_i \in \Sigma$ is defined by 
$\height(\alpha) = \sum_{i=1}^n c_i$.
For a nonzero integer $k$, let $\g_k$ be the span of root vectors 
$X_\alpha$ with $\height(\alpha) = k$.
And we put $\g_0 = \h$.
Then we have
\[
\g = \bigoplus_{k \in \Int} \g_k,
\quad
[\g_k, \g_l] \subset \g_{k+l}.
$$
If we put
$$
L_1 = \sum_{i=1}^n X_{\alpha_i},
\quad
L_0 = \sum_{i=1}^n a_i(t) H_i,
\quad
L_{-k} = \sum_{\height(\alpha) = k} b_\alpha(t) X_{-\alpha}
\qquad(k > 0),
\]
then we have
\[
L = L_1 + L_0 + L_{-1} + L_{-2} + \cdots,
\quad
L^{\ge 0} = L_1 + L_0,
\quad
L_k \in \g_k.
\]
By comparing the graded components of \eqref{eq:fKT1}, we obtain
\begin{gather}
\label{eq:deg0}
\frac{ \partial L_0 }{\partial t_1 } = [L_1, L_{-1}],
\\
\label{eq:degk}
\frac{ \partial L_{-k} }{\partial t_1 } = [L_1,L_{-k+1}] + [L_0, L_{-k}]
\quad(k>0).
\end{gather}

Since $[L_1, L_{-1}] = \sum_{i=1}^n b_{\alpha_i} H_i$, we obtain \eqref{eq:b=a'} 
from \eqref{eq:deg0} by comparing the coefficients of $H_i$.

We prove by induction on $\height(\beta)$ that 
$b_\beta$ can be uniquely expressed as a linear combination 
of $a_i b_\alpha$ and $\frac{\partial b_\alpha}{\partial t_1}$ with 
$1 \le i \le n$ and $\height(\alpha) = \height(\beta)-1$.
Suppose $k>0$.
Equating the coefficients of $X_{-\alpha}$ with $\height(\alpha) = k$ 
in \eqref{eq:degk}, we obtain
\begin{equation}
\label{eq:lin_eq}
\frac{\partial b_\alpha}{\partial t_1}
 =
- \left(\sum_{i=1}^n\alpha(H_i) a_i\right) b_\alpha 
+ \sum_{\height(\beta)=k+1} N_{\alpha-\beta,\beta} b_\beta
\qquad(\height(\alpha) = k),
\end{equation}
where $N_{\alpha-\beta,\beta} = 0$ unless $\alpha - \beta$ is a simple root.
We regard \eqref{eq:lin_eq} as a system of linear equations in the unknown variables 
$b_\beta$ ($\height(\beta)=k+1$) and consider the coefficient matrix
$$
M
 = 
\left( N_{\alpha - \beta,\beta} \right)_{\height(\alpha) = k, \, \height(\beta)=k+1}.
$$
On the other hand, it can be shown that $M$ is the representation matrix 
of the linear map $\ad L_1 : \g_{-(k+1)} \to \g_{-k}$.
Since $L_1 = \sum_{i=1}^n X_{\alpha_i}$, 
we can find a principal $\spl_2$-triple $\{ h, e, f \}$ 
with $e = L_1$ such that $[h,e] = 2e$, $[h,f] = -2f$ and $[e,f] = h$, 
and we have $\g_k = \{ X \in \g : [h,X] = 2k X \}$ (see \cite[Section~5]{Kos:59}).
By appealing to the representation theory of $\spl_2$, 
we see that $\ad L_1 : \g_{-(k+1)} \to \dim \g_{-k}$ is injective.
Hence the matrix $M$ has a full rank 
and the system \eqref{eq:lin_eq} of linear equations has a unique solution 
in $(b_\beta)_{\height(\beta) = k+1}$.
\end{proof}

\begin{remark}
Noting \eqref{eq:lin_eq}, 
one can impose the following constraints (or reduction, see \cite[Section~5]{KY:96}):
\[
b_\alpha = 0
\quad\text{with}\quad \height(\alpha) \ge k+1.
\]
The resulting hierarchy may be called the \emph{$k$-banded} f-KT hierarchy. 
For example, the KT hierarchy for a tridiagonal Lax matrix is the $1$-banded f-KT hierarchy. 
Then the constraints lead to additional equations for the $\tau$-functions 
(see Remark~\ref{rem:KT-ab}).
\end{remark}

We can find a solution of \eqref{eq:fKT} by considering the Gauss decomposition 
of $\exp \Theta(L(0);\tv)$ (this is a standard method to find the solution, 
see e.g. \cite{GW:84, KW:15}).
In general we have a Bruhat decomposition
$$
\exp \Theta(L(0);\tv) = \overline{n}(\tv) \dot{w} b(\tv)
$$
with $\overline{n}(\tv) \in \overline{\NN}$, $w \in W$ and $b(\tv) \in \BB$ for a particular $\tv$, 
where $\dot{w}$ is a fixed representative of the coset $w \in W = N_{\GG}(\HH)/\HH$.
It can be shown that $w = e$ if and only if $\tau_k(\tv) \neq 0$ for all $1 \le k \le n$ 
(see e.g. \cite{KW:15}). 

\begin{proposition}
\label{prop:solution}
For a generic $\tv$ (i.e. $\tau_k(\tv) \neq 0$ for $1 \le k \le n$), 
we have the Gauss decomposition
\begin{equation}
\label{eq:Gauss}
\exp \Theta(L(0);\tv) = \overline{n}(\tv) b(\tv),
\end{equation}
with $\overline{n}(\tv) \in \overline{\NN}$ and $b(\tv) \in \BB$, 
and the solution of the f-KT hierarchy \eqref{eq:fKT} is given by
\begin{equation}
\label{eq:solution}
L(\tv) = \Ad (\overline{n}(\tv)^{-1}) L(0) = \Ad (b(\tv)) L(0).
\end{equation}
Moreover, we have
\begin{equation}
\label{eq:a=tau}
a_i(\tv)
 =
\frac{\partial}{\partial t_1} \ln \tau_i(L(0);\tv)
 =
\frac{ \tau_i'(L(0);\tv) }{ \tau_i(L(0);\tv) }
\qquad(1 \le i \le n),
\end{equation}
where $\tau_i' = \frac{\partial \tau_i}{\partial t_1}$, 
and $b_\alpha(\tv)$ are expressed as rational functions 
of $\tau_i(L(0);\tv)$ and their derivatives.
\end{proposition}

\begin{remark}
\label{rem:KT-ab}
In the case of the KT lattice, one can find the following formulas 
of the solutions in terms of $\tau$-functions (see also \cite{GW:84, FH:91}):
\[
a_i(\tv)
 =
\frac{\partial}{\partial t_1} \ln \tau_i(\tv),
\quad
b_{\alpha_i}(\tv)
 =
b_{\alpha_i}(0) \prod_{j=1}^n \tau_j(\tv)^{-C_{i,j}}
\qquad (1 \le i \le n).
\]
Note that the $\tau$-functions for the KT hierarchy satisfy
\[
\frac{\partial^2}{\partial t_1^2} \ln \, \tau_i(\tv)
 =
b_{\alpha_i}(0) \prod_{j=1}^n \tau_j(\tv)^{-C_{i,j}}
\qquad (1 \le i \le n).
\]
\end{remark}

\begin{proof}[Proof of Proposition~\ref{prop:solution}]
We may assume that $\GG$ is a matrix group.

By differentiating the both sides of \eqref{eq:Gauss} with respect to $t_k$, we obtain
$$
P_k(0) \cdot \exp \Theta(L(0);\tv) = \exp \Theta(L(0);\tv) \cdot P_k(0)
 =
\frac{\partial \overline{n}}{\partial t_k} b
 +
\overline{n} \frac{\partial b}{\partial t_k},
$$
where $P_k(0) = (\nabla I_k)(L(0))$.
Multiplying this identity on the left by $\overline{n}^{-1}$ and on the right by $b^{-1}$ 
yields 
\begin{equation}
\label{eq:Pk}
\overline{n}^{-1} P_k(0) \overline{n}
 =
b P_k(0) b^{-1}
 =
\overline{n}^{-1} \frac{\partial \overline{n}}{\partial t_k}
 +
\frac{\partial b}{\partial t_k} b^{-1}.
\end{equation}
Here we recall that $P_1(\tv) = (\nabla I_1)(L(\tv)) = L(\tv)$.

We put
$$
\tilde{P}_k(\tv)
 = 
\overline{n}^{-1}(\tv) P_k(0) \overline{n}(\tv)
 =
b(\tv) P_k(0) b(\tv)^{-1},
\quad
\tilde{L}(\tv) = \tilde{P}_1(\tv),
$$
and prove $\tilde{L}(\tv) = L(\tv)$.
Since $[\overline{\n},L(0)] \in \overline{\bb}$, 
we see that $\tilde{L}(\tv)$ has the form \eqref{eq:Lax}.

First we show that $\tilde{L}(\tv)$ solves the f-KT hierarchy \eqref{eq:fKT}.
By using \eqref{eq:grad1}, we obtain
$$
\tilde{P}_k(\tv)
 =
\Ad(b(\tv)) P_k(0)
 = 
\Ad(b(\tv)) (\nabla I_k)(L(0))
 =
(\nabla I_k) (\Ad(b(\tv)) L(0) )
 =
(\nabla I_k) (\tilde{L}(\tv)).
$$
Also it follows from \eqref{eq:Pk} that
$$
\tilde{P}_k^{\ge 0}
 =
\frac{\partial b}{\partial t_k} b^{-1}.
$$
Now we prove
$$
\frac{\partial \tilde{L}}{\partial t_k}
 =
[ \tilde{P}_k^{\ge 0}, \tilde{L} ].
$$
Differentiating $\tilde{L}(\tv) = b(\tv) L(0) b(\tv)^{-1}$ with respect to $t_k$, 
we have
\begin{align*}
\frac{\partial \tilde{L}}{\partial t_k}
 &=
\frac{\partial b}{\partial t_k} L(0) b^{-1}
 +
b L(0) \cdot \left( - b^{-1} \frac{\partial b}{\partial t_k} b^{-1} \right)
 =
\frac{\partial b}{\partial t_k} b^{-1} \cdot b L(0) b^{-1}
 -
b L(0) b^{-1} \cdot \frac{\partial b}{\partial t_k} b^{-1}
\\
 &=
[ \tilde{P}_k^{\ge 0}, \tilde{L} ].
\end{align*}
Therefore, since $\tilde{L}(0) = L(0)$, 
we have $\tilde{L}(\tv) = L(\tv)$ near the origin $\tv = 0$ 
by the uniqueness theorem for differential equations.

Next we show that $L(\tv)$ is described in terms of the $\tau$-functions.
We decompose $b(\tv) = h(\tv) n(\tv)$ with $h(\tv) \in \HH$ and $n(\tv) \in \NN$.
Then we have
$$
L^{\ge 0}
 =
\frac{\partial b}{\partial t_1} b^{-1}
 =
\frac{\partial h}{\partial t_1} h^{-1}
 +
h \frac{\partial n}{\partial t_1} n^{-1} h^{-1}.
$$
Since $\frac{\partial h}{\partial t_1} h^{-1} \in \h$ and 
$h \frac{\partial n}{\partial t_1} n^{-1} h^{-1} \in \n$, 
we obtain
$$
\sum_{i=1}^n a_i H_i
 =
\frac{\partial h}{\partial t_1} h^{-1}.
$$
If $h(\tv) = \exp \left( \sum_{j=1}^n c_j(\tv) H_i \right)$, then we can see
$$
a_i(\tv) = \frac{\partial c_i(\tv)}{\partial t_1}.
$$
On the other hand, by using \eqref{eq:mat_coeff2} and 
$\varpi_i \left( \sum_{j=1}^n c_j(\tv) H_j \right) = c_i(\tv)$, 
we have
\[
\tau_i(\tv)
 = 
c_{\varpi_i}(\exp \Theta(L(0);\tv))
 = 
c_{\varpi_i}(\overline{n}(\tv) h(\tv) n(\tv))
 =
\chi^{\varpi_i}(h(\tv))
 =
e^{c_i(\tv)},
\]
thus we obtain $c_i(\tv) = \ln \tau_i(\tv)$.
Hence, by using Proposition~\ref{prop:sol_b}, we see that 
$b_\alpha(\tv)$s are determined by the $\tau$-functions.

Combining the above argument, we conclude $\tilde{L}(\tv) = L(\tv)$ for a generic $\tv$, 
which completes the proof.
\end{proof}

It is not always the case that $\exp \Theta(L(0);\tv)$ has the decomposition 
\eqref{eq:Gauss}.
For the case $\tau_k(L(0);\tv) = 0$ at some $\tv = \tv_*$ for some $k$, 
we have the following proposition.

\begin{proposition}
\label{prop:nongeneric}
Suppose that at $\tv = \tv_*$ we have a decomposition
\begin{equation}
\label{eq:Bruhat}
\exp \Theta(L(0);\tv_*) = \overline{n}_* \dot{w}_* h_* n_*
\end{equation}
with $\overline{n}_* \in \overline{\NN}$, $w_* \in W$, $h_* \in \HH$ and $n_* \in \NN$.
Then we have
\begin{equation}
\tau_k(L(0);\tv + \tv_*)
 =
\chi^{\varpi_k}(h_*)
\sum_{i=1}^r
 \langle u_i, \overline{n}_* \dot{w} \cdot v_{\varpi_k} \rangle_{\varpi_k}
 \tau_k(L(0);\tv;u_i),
\end{equation}
where $\{ u_1, \dots, u_r \}$ is an orthonormal weight basis of $V(\varpi_k)$.
\end{proposition}

We remark that $\langle u_i, \overline{n}_* \dot{w}\cdot v_{\varpi_k} \rangle_{\varpi_k} = 0$ 
unless the weight $\wt u_i$ of $u_i$ satisfies $\wt u_i \le w \varpi_k$, 
where we write $\lambda \le \mu$ if $\mu - \lambda$ is a linear combination of simple roots 
with nonnegative integer coefficients.

\begin{proof}
Since $n_*\cdot v_{\varpi_k} = v_{\varpi_k}$ and 
$h_*\cdot v_{\varpi_k} = \chi^{\varpi_k}(h_*) v_{\varpi_k}$, 
we have
\begin{align*}
\tau_k(L(0);\tv + \tv_*)
 &=
\langle
 v_{\varpi_k}, 
 \exp \Theta(L(0);\tv) \exp \Theta(L(0);\tv_*) \cdot v_{\varpi_k}
\rangle_{\varpi_k}
\\
 &=
\chi^{\varpi_k}(h_*)
\langle
 v_{\varpi_k},
 \exp \Theta(L(0);\tv) \overline{n}_* \dot{w}\cdot  v_{\varpi_k}
\rangle_{\varpi_k}.
\end{align*}
Now the desired identity is obtained by using \eqref{eq:mat_coeff1}.
\end{proof}

In this article we are interested in rational solutions of the f-KT hierarchy \eqref{eq:fKT}.
By Proposition~\ref{prop:solution}, it is enough to consider 
the case where $\tau_k(L(0);\tv)$ are polynomials in $\tv$,
i.e., $L(0)$ is nilpotent.
Let 
\[
\Lambda = \sum_{i=1}^n X_{\alpha_i}
\]
be the standard regular nilpotent element. 
Then we have the following proposition for the formula of polynomial $\tau$-functions.

\begin{proposition}
\label{prop:nilpotent}
If $L(0)$ is nilpotent, then there exists $\overline{n} \in \overline{\NN}$ 
such that $L(0) = \Ad(\overline{n}) \Lambda$, and we have
$$
\tau_k(L(0);\tv)
 =
\sum_{i=1}^r
 \langle u_i, \overline{n}^{-1} \cdot v_{\varpi_n} \rangle_{\varpi_k} 
 \tau_k(\Lambda;\tv;u_i)\qquad (1\le k\le n),
$$
where $\{ u_1, \dots, u_r \}$ is an orthonormal weight basis of $V(\varpi_k)$.
\end{proposition}

\begin{proof}
We note that $L(0) \in \Lambda + \overline{\bb}$.
It is known (see \cite[Proposition~16]{Kos:63}) that 
$X \in \g$ is nilpotent if and only if $I_1(X) = \dots = I_n(X) = 0$.
By \cite[Proposition~2.3.2]{Kos:79},  
the following are equivalent for $X$, $Y \in \Lambda + \overline{\bb}$:
\begin{enumerate}
\item[(i)]
there is an element $\overline{n} \in \overline{\NN}$ such that $\Ad(\overline{n}) X = Y$;
\item[(ii)]
$I_1(X) = I_1(Y)$, $\cdots$, $I_n(X) = I_n(Y)$.
\end{enumerate}
Hence we conclude that there exists an element $\overline{n} \in \overline{\NN}$ 
such that $L(0) = \Ad(\overline{n}) \Lambda$.

By applying \eqref{eq:grad1}, 
we have $\Theta(L(0);\tv) = \Ad(\overline{n}) \Theta(\Lambda;\tv)$, 
and $\exp \Theta(L(0);\tv)
 = \overline{n} \left( \exp \Theta(\Lambda;\tv) \right) \overline{n}^{-1}$.
Hence, by using $\theta(\overline{n})^{-1}\cdot v_{\varpi_k} = v_{\varpi_k}$ 
and \eqref{eq:mat_coeff2}, we have
\begin{align*}
\tau_k(L(0);\tv)
 &=
\langle
 v_{\varpi_k},
 \overline{n} \left( \exp \Theta(\Lambda;\tv) \right) \overline{n}^{-1} \cdot v_{\varpi_k}
\rangle_{\varpi_k}
 =
\langle
 v_{\varpi_k},
 \left( \exp \Theta(\Lambda;\tv) \right) \overline{n}^{-1}\cdot v_{\varpi_k}
\rangle_{\varpi_k}
\\
 &=
\sum_{i=1}^r
 \langle u_i, \overline{n}^{-1} \cdot v_{\varpi_k} \rangle_{\varpi_k} 
 \tau_k(\Lambda;\tv;u_i).
\end{align*}
\end{proof}

Proposition~\ref{prop:nilpotent} implies that, 
by considering all elements $\overline{n} \in \overline{\NN}$ 
and an orthonormal weight basis $\{ u_1, \dots, u_r \}$, 
one can obtain all the polynomial $\tau$-functions for the f-KT hierarchy 
on the Lie algebra $\g$.

\section{%
Matrix realization of the f-KT hierarchy for type $D$
}
\label{sec:fKT_D}

In this section we give an explicit matrix realization 
of the f-KT hierarchy \eqref{eq:fKT} of type $D_n$.

\subsection{%
Lie algebra of type $D$
}

We recall several basic facts on even orthogonal Lie algebras.

We use the following realization of the orthogonal Lie algebra $\sorth_{2n}(\Comp)$.
Let $S$ be the $2n \times 2n$ antidiagonal symmetric matrix given by
\footnote{%
The reason we choose this matrix $S$ is that the standard regular nilpotent element $\Lambda$
can be taken as a $(0,1)$-matrix. See \eqref{eq:reg_nilp_D}.}
\[
S
 = 
\sum_{i=1}^n (-1)^{n-i} ( E_{i,2n+1-i} + E_{2n+1-i,i} ),
\]
where $E_{i,j}$ denote the matrix unit with 1 at the $(i,j)$ entry 
and 0 at all other entries.
Then we define the orthogonal Lie algebra $\sorth_{2n}(\Comp)$, 
the simple Lie algebra of type $D_n$, 
by putting
\[
\sorth_{2n}(\Comp) = \{ X \in \gl_{2n}(\Comp) : \trans X S + S X = 0 \}.
\]
Then the corresponding orthogonal and special orthogonal groups are given by 
\[
\Orth_{2n}(\Comp) = \{ g \in \GL_{2n}(\Comp) : \trans g S g = S \},
\quad
\SOrth_{2n}(\Comp) = \{ g \in \SL_{2n}(\Comp) : \trans g S g = S \}
\]
respectively.
We denote by $\Spin_{2n}(\Comp)$ the spin group, which is the simply connected 
Lie group of type $D_n$.

In the remaining of this paper we write $\g = \sorth_{2n}(\Comp)$.
Let $\h$ (resp. $\n$, $\overline{\n}$) be a subalgebra of $\g$ 
consisting of diagonal matrices (resp. strictly upper triangular matrices, 
strictly lower triangular matrices) in $\g$.
Then 
$$
\h
 = 
\{ \diag (h_1, \dots, h_n, - h_n, \dots, -h_1) : h_i \in \Comp \}.
$$
is a Cartan subalgebra of $\g$ and 
$\g = \overline{n} \oplus \h \oplus \n$ is a triangular decomposition.
Let $\ep_i : \h \to \Comp$ be the linear map given by 
$\ep_i ( \diag (h_1, \dots, h_n, - h_n, \dots, -h_1) ) = h_i$.
Then $\{ \ep_i : 1 \le i \le n \}$ forms a basis of $\h^*$ 
and the root system of $\g$ with respect to $\h$ is given by
$$
\Sigma = \{ \pm (\ep_i - \ep_j),\, \pm (\ep_i + \ep_j) : 1 \le i < j \le n \}.
$$
We put
\begin{align*}
H_i
 &= 
E_{i,i} - E_{i+1,i+1} + E_{2n-i,2n-i} - E_{2n-i+1,2n-i+1}
\qquad(1 \le i \le n-1),
\\
H_n
 &= 
E_{n-1,n-1} - E_{n+1,n+1} + E_{n,n} - E_{n+2,n+2},
\end{align*}
and
\begin{equation*}
\begin{array}{lll}
X_{\ep_i - \ep_j}
 =
E_{i,j} - (-1)^{j-i} E_{2n+1-j,2n+1-i},
\\[1.0ex]
X_{\ep_i + \ep_j}
 =
E_{i,2n+1-j} - (-1)^{j-i} E_{j,2n+1-i},
\\[1.0ex]
X_{-(\ep_i - \ep_j)}
 =
E_{j,i} - (-1)^{j-i} E_{2n+1-i,2n+1-j},
\\[1.0ex]
X_{-(\ep_i + \ep_j)}
 =
E_{2n+1-j,i} - (-1)^{j-i} E_{2n+1-i,j},\\[0.5ex]
\end{array}
\qquad (1 \le i < j \le n).
\end{equation*}
These elements $\{ H_i : 1 \le i \le n \} \cup \{ X_\alpha : \alpha \in \Sigma \}$ 
form a Chevalley basis of $\g$.
The roots
\begin{equation}
\label{eq:simple}
\alpha_1 = \ep_1 - \ep_2,
\quad \dots,\quad
\alpha_{n-1} = \ep_{n-1} - \ep_n,
\quad
\alpha_n = \ep_{n-1} + \ep_n
\end{equation}
form the simple system corresponding to $\n$.
Moreover the bilinear form $\kappa : \g \times \g \to \Comp$ given by 
\begin{equation}
\label{eq:inv_form}
\kappa (X,Y) = \tr (XY)
\qquad
(X, Y \in \g)
\end{equation}
is symmetric, invariant and nondegenerate.

In this setting, the Lax matrix for $\sorth_8(\Comp)$ ($n=4$) 
given by \eqref{eq:Lax} has th following form:
\[
L
 =
\begin{pmatrix}
 a_1 & 1 &  &  &  &   &  & \\
 b_{\ep_1 - \ep_2} & a_2-a_1 & 1 & & & & & \\
 b_{\ep_1 - \ep_3} & b_{\ep_2 - \ep_3} & a_3-a_2+a_4 & 1 & 1& &  & \\
 b_{\ep_1 - \ep_4} & b_{\ep_2 - \ep_4} & b_{\ep_3 - \ep_4} & a_4-a_3 & 0 & 1 & & \\
 b_{\ep_1 + \ep_4} & b_{\ep_2 + \ep_4} & b_{\ep_3 + \ep_4} & 0 & a_3-a_4 & 1 & & \\
 b_{\ep_1 + \ep_3} & b_{\ep_2 + \ep_3} & 0 & b_{\ep_3+\ep_4} & b_{\ep_e - \ep_4} & a_2-a_3-a_4 & 1 & \\
 b_{\ep_1 + \ep_2} & 0 & b_{\ep_2 + \ep_3} & -b_{\ep_2+\ep_4}
 & b_{\ep_2 - \ep_4} & - b_{\ep_2 - \ep_3} & a_1-a_2& 1\\
 0 & b_{\ep_1 + \ep_2} & - b_{\ep_1 + \ep_3} & b_{\ep_1 + \ep_4} 
 & b_{\ep_1 - \ep_4} & - b_{\ep_1-\ep_3} & b_{\ep_1 - \ep_2}& -a_1
\end{pmatrix},
\]
where all unfilled entries in the upper triangular part are zero. 

Let $W$ be the Weyl group of $\g = \sorth_{2n}(\Comp)$.
Then $W$ acts on the Cartan subalgebra $\h$ as permutations 
together with an even number of sign changes in the coordinates 
$h_1, \dots, h_n$.
Hence $W$ is isomorphic to the subgroup of the symmetric group $\Sym_{2n}$ 
given by
\begin{equation}
\label{eq:Weyl}
\WD_n
 =
\left\{ 
 w \in \Sym_{2n} :
 \begin{array}{l}
  \text{$w(i) + w(2n+1-i) = 2n+1$ for $1 \le i \le 2n$} \\
  \text{$\# \{ i : 1 \le i \le n, \, w(i) \ge n+1 \}$ is even}
 \end{array}
\right\},
\end{equation}
and the isomorphism is given by
\[
s_i \mapsto (i,i+1)(2n-i,2n+1-i)
\quad(1 \le i \le n-1),
\qquad
s_n \mapsto (n-1,n+1)(n,n+2),
\]
where $s_i$ is the simple reflection corresponding to $\alpha_i$ 
and $(p,q) \in \Sym_{2n}$ denotes the transposition of $p$ and $q$.

\subsection{%
Explicit form of the f-KT hierarchy on $\sorth_{2n}(\Comp)$
}

Let $\g = \sorth_{2n}(\Comp)$ and $\GG = \SOrth_{2n}(\Comp)$ or $\Spin_{2n}(\Comp)$.
First we give an explicit system of algebraically independent generators of 
the ring $\Comp[\g]^{\GG}$ of invariant polynomial functions on $\g$.
We denote by $\Pf(A)$ the Pfaffian of a skew-symmetric matrix $A$ of even order.
For the invariants of the ring of polynomials $\Comp[\g]$, we have the following proposition.

\begin{proposition}
\label{prop:Chevalley}
We define $I_k \in \Comp[\g]$ ($1 \le k \le n$) by putting
\begin{equation}
\label{eq:Chevalley}
I_k(X) = \frac{1}{2k} \tr (X^{2k})
\quad
(1 \le k \le n-1),
\qquad
I_n(X) = 2 c_n \Pf (SX),
\end{equation}
where $c_n = -(-1)^{n(n+1)/2}$ is a normalization constant (see \eqref{eq:norm}).
Then $\{ I_1, \dots, I_{n-1}, I_n \}$ forms a system of algebraically independent 
homogeneous generators of $\Comp[\g]^{\GG}$.
\end{proposition}

\begin{proof}
Let $R : \Comp[\g] \to \Comp[\h]$ be the restriction map.
Then the Chevalley restriction theorem asserts that $R$ induces 
an isomorphism $\Comp[\g]^{\GG} \to \Comp[\h]^W$. 
If we identify $\Comp[\h]$ with the polynomial algebra in $(x_1, \dots, x_n)$, 
where $x_i$ corresponds to $\ep_i \in \h^*$, then we have
$$
R(I_k) = \frac{1}{k} (x_1^{2k} + \dots + x_n^{2k})
\quad(1 \le k \le n-1),
\qquad
R(I_n)
 = 
-2 x_1 x_2 \dots x_n.
$$
Also it is known (see \cite[p.68]{Hum:90}) 
that $\Comp[\h]^W$ is generated by 
$n$ algebraically independent homogeneous polynomials
$$
f_k = x_1^{2k} + \dots + x_n^{2k}
\quad(1 \le k \le n-1),
\qquad
f_n = x_1 x_2 \cdots x_n.
$$
The proposition follows from these claims.
\end{proof}

Given a skew-symmetric matrix $A$ of even order, 
let $\hat{A}$ is the skew-symmetric matrix, called the \emph{coPfaffian matrix} of $A$, 
whose $(i,j)$ entry, $i<j$, is given by
\begin{equation}
\label{eq:coPf}
\hat{A}_{i,j} = (-1)^{i+j+1} \Pf (A^{i,j}),
\end{equation}
where $A^{i,j}$ denotes the submatrix of $A$ obtained from $A$ 
by removing the $i$th and $j$th rows/columns.
Then we have
$$
A \cdot \trans\hat{A} = \trans\hat{A} \cdot A = \Pf (A) \cdot I,
$$
where $I$ is the identity matrix.
(We refer the readers to \cite{IO:14} for an exposition on Pfaffians.)
The gradients of $I_k~(1\le k\le n)$  are given as follows.

\begin{proposition}
\label{prop:grad_D}
We have
\begin{equation}
\label{eq:grad_D}
\left( \nabla I_k \right)(X) = X^{2k-1}\quad (1\le k\le n-1),
\qquad
\left( \nabla I_n \right)(X) =c_n \trans \hat{SX} \cdot S.
\end{equation}
\end{proposition}

In the proof of the second identity, we need the following formula 
for the derivative of a Pfaffian.

\begin{lemma}
\label{lem:dPf}
Let $n$ be an even integer.
If $A(t) = \left( a_{i,j}(t) \right)_{1 \le i, j \le n}$ is a skew-symmetric matrix 
with entries functions of $t$, 
then we have
\begin{equation}
\label{eq:dPf}
\frac{d}{dt} \Pf( A(t))
 =
\frac{1}{2} \tr \left( \trans \hat{A(t)} \cdot A'(t) \right),
\end{equation}
where $A'(t) = \left( \frac{d}{dt} a_{i,j}(t) \right)_{1 \le i, j \le n}$.
\end{lemma}

\begin{proof}
If we regard the Pfaffian $\Pf(A)$ as a function in its entries $a_{i,j}$ ($i<j$), 
then it follows from the expansion of $\Pf(A)$ along the $i$th row that
$$
\frac{\partial}{\partial a_{i,j}} \Pf (A)
 =
(-1)^{i+j+1} \Pf (A^{i,j})
 =
\hat{A}_{i,j}.
$$
Hence we see that
$$
\frac{d}{dt} \Pf (A(t))
 =
\sum_{1 \le i < j \le n}
 \frac{\partial}{\partial a_{i,j}} \Pf (A)
 \cdot
 \frac{d}{dt} a_{i,j}
 =
\sum_{1 \le i < j \le n}
 \hat{A}_{i,j}
 \frac{d}{dt} a_{i,j}.
$$
Since $\hat{A}_{i,i} = 0$, $\hat{A}_{i,j} = - \hat{A}_{j,i}$, 
and $A'_{i,j} = - A'_{j,i}$, we have
$$
\frac{1}{2} \tr \left( \trans \hat{A(t)} \cdot A'(t) \right)
 =
\frac{1}{2} \sum_{i,j=1}^n \hat{A}_{i,j} A'_{i,j}
 =
\sum_{1 \le i < j \le n} \hat{A}_{i,j} A'_{i,j}.
$$
This completes the proof.
\end{proof}

\begin{proof}[Proof of Proposition~\ref{prop:grad_D}]
The proof for the first identity is easy.
We prove the second identity.
If we put $Z(t) = S(X+tY)$, then we have
$$
\kappa \left( \nabla I_n)(X), Y \right)
 =
\frac{d}{dt} I_n(X+tY) \big|_{t=0}
 =
2c_n \frac{d}{dt} \Pf (Z(t)) \big|_{t=0}.
$$
By using Lemma~\ref{lem:dPf}, we have
$$
\kappa \left( (\nabla I_n)(X), Y \right)
 =
c_n\tr \left( \trans \hat{Z} \cdot Z'(t) \right) \big|_{t=0}
 =
c_n\tr \left( \trans \hat{SX} \cdot SY \right)
 =
c_n\kappa \left( \trans \hat{SX} \cdot S, Y \right).
$$
Since $\kappa$ is nondegenerate on $\g$, 
we obtain the desired identity.
\end{proof}

We rename the time variables from $(t_1, \dots, t_n)$ 
to $(t_1, t_3, \dots, t_{2n-3}, s)$, 
where $t_{2k-1}$ corresponds to the trace invariant $I_k(X)$
and $s$ corresponds to the Pfaffian invariant $I_n(X)$.  In summary, we obtain the following proposition.

\begin{proposition}
\label{prop:fKT_D}
The matrix representation of the f-KT hierarchy 
on the Lie algebra $\sorth_{2n}(\Comp)$ is given as follows:
\begin{equation}
\label{eq:fKT_D}
\begin{cases}
 \dfrac{\partial L}{\partial t_{2k-1}}
  = 
 \left[
  \left( L^{2k-1} \right)^{\ge 0}, L
 \right]
 \qquad (1 \le k \le n-1),
\\[10pt]
\quad \dfrac{\partial L}{\partial s}
 =
c_n \left[
  \big(\trans \hat{SL} \cdot S \big)^{\ge 0}, L
 \right],
\end{cases}
\end{equation}
where $c_n = -(-1)^{n(n+1)/2}$.
\end{proposition}

\section{%
Polynomial solutions of the $\tau$-functions for type $D$
}\label{sec:Psolutions}

In this section, we give explicit formulas for the polynomial $\tau$-functions 
for type $D$.
By Propositions~\ref{prop:nongeneric} and \ref{prop:nilpotent}, 
it is enough to consider 
the $\tau$-functions $\tau_r(\Lambda;\tv;u)$ associated to 
the standard regular nilpotent element $\Lambda$.

\subsection{%
Explicit formulas for $\tau_n$ and $\tau_{n-1}$
}

Recall the definition of the $\tau$-functions in the case of the orthogonal Lie algebra 
$\g = \sorth_{2n}(\Comp)$ of type $D_n$.
Let $\tv = (t_1, \dots, t_{2n-3}, s)$ be the time variables of the f-KT hierarchy 
(\ref{eq:fKT_D}) for $\g$. 
Let
\begin{equation}
\label{eq:reg_nilp_D}
\Lambda
 = 
\sum_{i=1}^n X_{\alpha_i}
 =
\sum_{i=1}^{n-1} \left( E_{i,i+1} + E_{2n-i,2n+1-i} \right)
 +
E_{n-1,n+1} + E_{n,n+2}
\end{equation}
be the standard regular nilpotent element of $\g$, 
and put
\[
\Theta(\Lambda ; \tv)
 = 
\sum_{k=1}^{n-1} t_{2k-1} \Lambda^{2k-1}
 +
s c_n\trans ( \hat{S \Lambda} ) \cdot S
\]
where $\hat{A}$ denotes the coPfaffian matrix of a skew-symmetric matrix $A$ 
(see \eqref{eq:coPf}).
The fundamental weights of $\g$ corresponding to the simple roots defined 
by \eqref{eq:simple} are given by 
\begin{equation}
\label{eq:fund_wt}
\begin{gathered}
\varpi_{i} = \ep_1 + \dots + \ep_{i}
\qquad (1\le i\le n-2),
\\
\varpi_{n-1} = \frac{1}{2} \left( \ep_1 + \dots + \ep_{n-2} + \ep_{n-1} - \ep_n \right),
\\
\varpi_n = \frac{1}{2} \left( \ep_1 + \dots + \ep_{n-2} + \ep_{n-1} + \ep_n \right).
\end{gathered}
\end{equation}
Then, from Definition \ref{def:tau}, the $\tau$-functions are defined as the matrix coefficients 
on the irreducible highest weight representations $V(\varpi_r)$ 
with highest weight $\varpi_r$ and highest weight vector $v_{\varpi_r}$:
$$
\tau_r(\Lambda;\tv;u)
 =
\langle v_{\varpi_r}, \exp \Theta(\Lambda;\tv) \cdot u \rangle_{\varpi_r},
$$
where $u \in V(\varpi_r)$ and 
$\langle \cdot, \cdot \rangle_{\varpi_r}$ is the bilinear form 
given in Lemma~\ref{lem:bilinear}.

To state our main results, we introduce $2$-reduced Schur functions and Schur's $Q$-functions 
in the so-called Sato-variables $\tv' = (t_1, t_3, \dots, t_{2n-3})$ 
(see \cite[III.8]{Mac:95} and \cite{Oka:19} for Schur's $Q$-functions).
We define $q_m(\tv')$ by the generating function
\begin{equation}
\label{eq:q}
\sum_{m \ge 0} q_m(\tv') z^m
 =
\exp \left( \sum_{k=1}^{n-1} t_{2k-1} z^{2k-1} \right).
\end{equation}
For a partition $\lambda = (\lambda_1, \dots, \lambda_n)$ 
($\lambda_1 \ge \cdots \ge \lambda_n \ge 0$) of length at most $n$, 
we define the \emph{$2$-reduced Schur function} $S_\lambda(\tv')$ by
\begin{equation}
\label{eq:Schur}
S_\lambda(\tv')
 = 
\det \left( q_{\lambda_i-i+j}(\tv') \right)_{1 \le i, j \le n}.
\end{equation}
For a strict partition $\alpha = (\alpha_1, \dots, \alpha_l)$ ($\alpha_1 > \dots > \alpha_l > 0$) 
of length $l$, 
the corresponding \emph{Schur's $Q$-function} $Q_\lambda(\tv')$ 
is defined inductively on $l$:
\begin{enumerate}
\item[(1)]
If $l=0$, i.e. $\lambda = \emptyset$, then $Q_\emptyset(\tv') = 1$.
\item[(2)]
If $l=1$, i.e. $\lambda = (i)$ with $i\ge 0$, then $Q_{(i)}(\tv') = q_i(\tv')$.
\item[(3)]
If $l =2$, i.e. $\lambda = (i,j)$ with $i>j>0$, then
\begin{equation}
\label{eq:Q2}
Q_{(i,j)}(\tv') = q_i(\tv') q_j(\tv') + 2 \sum_{k=1}^j (-1)^k q_{i+k}(\tv') q_{j-k}(\tv').
\end{equation}
\item[(4)]
If $l \ge 3$, then we define 
\begin{equation}
\label{eq:Q-Pf}
Q_\alpha(\tv')
 = 
\begin{cases}
 \Pf \big( Q_{(\alpha_i, \alpha_j)}(\tv') \big)_{1 \le i < j \le l}
 &\text{if $l$ is even,} \\
 \Pf \big( Q_{(\alpha_i, \alpha_j)}(\tv') \big)_{1 \le i < j \le l+1}
 &\text{if $l$ is odd,} \\
\end{cases}
\end{equation}
where $\alpha_{l+1} = 0$ and $Q_{(\alpha_i,0)}(\tv') = q_{\alpha_i}(\tv')$ when $l$ is odd.
\end{enumerate}

For a strict partition $\alpha$ such that $\alpha_1 \le n-1$, 
we define an \emph{extended Schur's $Q$-function} by
\begin{equation}
\label{eq:QD}
\hat{Q}_\alpha (\tv',s)
 =
\begin{cases}
 Q_\alpha(\tv') + (-1)^n s Q_{\alpha \setminus (n-1)}(\tv')
 &\text{if $n-1$ is a part of $\alpha$,} \\
 Q_\alpha(\tv')
 &\text{otherwise}
\end{cases}
\end{equation}
where $\alpha \setminus (n-1)$ is the strict partition obtained from $\alpha$ 
by removing the part $n-1$, 
i.e. the shifted diagram of $\alpha \setminus (n-1)$ is the same as the skew diagram $\alpha/(n-1)$.

Let $w \in W$ be a Weyl group element and regard it as a permutation in $\WD_n$ 
(see \eqref{eq:Weyl}).
Then we associate a strict partition $\alpha = \alpha(w) = (\alpha_1, \dots, \alpha_p)$ 
determined by the relation
\begin{equation}
\label{eq:alpha}
\{ \alpha_1 + n+1, \dots, \alpha_p + n+1 \}
 =
\{ w(i) : 1 \le i \le n, \, w(i) \ge n+2 \}.
\end{equation}

One of the main results in this paper is the following explicit formula 
of the $\tau$-functions $\tau_n$ and $\tau_{n-1}$.
(See Proposition~\ref{prop:tau} for the explicit formulas for other 
$\tau$-functions $\tau_r$ with $1 \le r \le n-2$.) 
Then the polynomial $\tau$-functions are given by the formulas in
Proposition \ref{prop:nilpotent}.

\begin{theorem}
\label{thm:tau}
For $w \in W \cong \WD_n$, we have
\begin{align}
\label{eq:tau=Q1}
\tau_n(\Lambda;\tv;\dot{w}\cdot v_{\varpi_n})
 &=
c_{n,w} \hat{Q}_{\alpha(w)}(\tv',s),
\\
\label{eq:tau=Q2}
\tau_{n-1}(\Lambda;\tv;\dot{w}\cdot v_{\varpi_{n-1}})
 &=
c_{n-1,w} \hat{Q}_{\alpha(w^\dagger)}(\tv',-s),
\end{align}
where $c_{n,w}$ and $c_{n-1,w}$ are nonzero constants, 
and $\WD_n \ni w \mapsto w^\dagger \in \WD_n$ is the automorphism given by 
$w^\dagger = (n,n+1) w (n,n+1)$. 
\end{theorem}

Before proving this theorem in the next subsection, we give a Pfaffian expression 
of $\hat{Q}_\alpha(\tv',s)$.

\begin{proposition}
For a strict partition $\alpha$ of length $l$ such that $\alpha_1 \le n-1$, we have
\begin{equation}
\label{eq:QD-Pf}
\hat{Q}_\alpha(\tv',s)
 =
\begin{cases}
 \Pf \left( \hat{Q}_{(\alpha_i,\alpha_j)} (\tv',s) \right)_{1 \le i < j \le l}
 &\text{if $l$ is even,}
\\
 \Pf \left( \hat{Q}_{(\alpha_i,\alpha_j)} (\tv',s) \right)_{1 \le i < j \le l+1}
 &\text{if $l$ is odd,}
\end{cases}
\end{equation}
where $\alpha_{l+1} = 0$ and $\hat{Q}_{(\alpha_i,0)} = \hat{Q}_{(\alpha_i)}$.
Moreover we have for $n-1 \ge i > j > 0$ 
\begin{equation}
\label{eq:Q2D}
\hat{Q}_{(i,j)}(\tv',s)
 = 
\hat{Q}_{(i)}(\tv',s) \hat{Q}_{(j)}(\tv',s)
 + 
2 \sum_{k=1}^j (-1)^k \hat{Q}_{(i+k)}(\tv',s) \hat{Q}_{(j-k)}(\tv',s),
\end{equation}
where $\hat{Q}_{(0)} = 1$.
\end{proposition}

\begin{proof}
Equation (\ref{eq:Q2D}) is an immediate consequence of (\ref{eq:Q2}) 
and the definition (\ref{eq:QD}).
For the proof of (\ref{eq:QD-Pf}), it is enough to consider the case where $\alpha_1 = n-1$.

If $l$ is even, then we use the multilinearity of Pfaffians to obtain 
$$
\Pf \left( \hat{Q}_{(\alpha_i,\alpha_j)}(\tv',s) \right)_{1 \le i < j \le l}
 =
\Pf \left( Q_{(\alpha_i,\alpha_j)}(\tv') \right)_{1 \le i < j \le l}
 +
(-1)^n s
\Pf \begin{pmatrix}
 0 & \left( Q_{(\alpha_i)}(\tv') \right)_{2 \le i \le l} \\
   & \left( Q_{(\alpha_i,\alpha_j)}(\tv') \right)_{2 \le i < j \le l}
\end{pmatrix}.
$$
The first Pfaffian is equal to $Q_\alpha(\tv')$.
By moving the first row/column to the last 
and then by multiplying the last row/column by $-1$, 
we see that the second Pfaffian equals 
$(-1)^{l-1} \cdot (-1) Q_{(\alpha_2, \dots, \alpha_l)}(\tv')
 = Q_{(\alpha_2, \dots, \alpha_l)}(\tv')$.

If $l$ is odd, then we have
\begin{multline*}
\Pf \left( \hat{Q}_{(\alpha_i,\alpha_j)}(\tv',s) \right)_{1 \le i < j \le l+1}
\\
 =
\Pf \left( Q_{(\alpha_i,\alpha_j)}(\tv') \right)_{1 \le i < j \le l+1}
 +
(-1)^n s
\Pf \begin{pmatrix}
 0 & \left( Q_{\alpha_i}(\tv') \right)_{2 \le i \le l} & 1 \\
   & \left( Q_{(\alpha_i,\alpha_j)}(\tv') \right)_{2 \le i < j \le l}
   & \left( Q_{\alpha_j}(\tv') \right)_{2 \le j \le l} \\
   &   & 0
\end{pmatrix}.
\end{multline*}
The first Pfaffian is equal to $Q_\alpha(\tv')$.
By adding the last row/column to the first row/column 
and then by expanding the resulting Pfaffian along the first row/column, 
we see that the second Pfaffian equals $Q_{(\alpha_2, \dots, \alpha_l)}(\tv')$.
\end{proof}

\subsection{%
Proof of Theorem~\ref{thm:tau}
}
\label{sec:proof}

As the first step toward the proof, we compute the explicit form of 
$\exp \Theta(\Lambda;\tv) \in \SOrth_{2n}(\Comp)$.

\begin{lemma}
\label{lem:expTheta}
If we put
\begin{align*}
\Lambda'
 &=
X_{\ep_1 + \ep_n} - X_{\ep_1 - \ep_n}
 =
-E_{1,n} + E_{1,n+1} + (-1)^n E_{n,2n} - (-1)^n E_{n+1,2n}
\\
\Lambda''
 &=
E_{1,2n}
\end{align*}
then we have
$$
\exp \Theta(\Lambda;\tv)
 =
\sum_{m=0}^{2n-2} q_m(\tv') \Lambda^m + s \Lambda' - (-1)^n s^2 \Lambda''.
$$
More explicitly, this matrix looks like
$$
\begin{pmatrix}
1 & q_1 & q_2    & \cdots  & q_{n-2} & q_{n-1} - s 
 & q_{n-1} + s & 2 q_n     & \cdots & 2 q_{2n-3} & 2 q_{2n-2} - (-1)^n s^2 \\
  & 1   & q_1    & \cdots  & q_{n-3} & q_{n-2} 
 & q_{n-2}     & 2 q_{n-1} & \cdots & 2q_{2n-4}  & 2q_{2n-3} \\
  &     & \ddots & \ddots  & \vdots  & \vdots 
 & \vdots      & \vdots    &        & \vdots     & \vdots \\
  &     &        & \ddots  & \vdots  & \vdots 
 & \vdots      & \vdots    &        & \vdots     & \vdots \\
 &      &        &         & 1       & q_1
 & q_1         & 2q_2      & \cdots & 2 q_{n-1}  & 2 q_n \\
 &      &        &         &         & 1 
 & 0           & q_1       & \cdots & q_{n-2}    & q_{n-1} + (-1)^n s \\
 &      &        &         &         & 
 & 1           & q_1       & \cdots & q_{n-2}    & q_{n-1} - (-1)^n s \\
 &      &        &         &         & 
 &             & 1         & \cdots & q_{n-3}    & q_{n-2} \\
 &      &        &         &         & 
 &             &           & \ddots & \vdots     & \vdots \\
 &      &        &         &         & 
 &             &           &        & 1          & q_1 \\
 &      &        &         &         & 
 &             &           &        &            & 1
 \end{pmatrix},
$$
where blank entries are all $0$.
\end{lemma}

\begin{proof}
First we prove
\begin{equation}
\label{eq:norm}
\trans \hat{S \Lambda} \cdot S
 = 
c_n \Lambda'
\qquad\text{with}\quad c_n = -(-1)^{n(n+1)/2}.
\end{equation}
Since the $n$th and $(n+1)$st rows/columns of $S \Lambda$ are the same 
and the $1$st row/column is the zero vector, 
we see that the $(i,j)$ entry, $i < j$, of $\hat{S \Lambda}$ vanishes 
unless $(i,j) = (1,n)$ or $(1,n+1)$.
The entries $\hat{S \Lambda}_{1,n}$ and $\hat{S \Lambda}_{1,n+1}$ 
can be obtained by a direct computation.

Since $\Lambda^{2n-1} = 0$, $\Lambda \Lambda' = \Lambda' \Lambda = 0$ and 
$(\Lambda')^2 =- 2 (-1)^n \Lambda''$, $(\Lambda')^3 = 0$, we have
$$
\exp \Theta (\Lambda;\tv)
 =
\exp \left( \sum_{k=1}^{n-1} t_{2k-1} \Lambda^{2k-1} \right)
\cdot
\exp \left( s \Lambda' \right)
 =
\left( \sum_{m=0}^{2n-2} q_m(\tv') \Lambda^m \right)
\cdot
\left( I + s \Lambda' - (-1)^n s^2 \Lambda'' \right).
$$
Now the explicit computation of $\Lambda^m$ completes the proof.
\end{proof}

The next step is to relate the $\tau$-functions with minors of 
$\exp \Theta(\Lambda;\tv)$.
Let $\Comp^{2n}$ be the defining representation of $\SOrth_{2n}(\Comp)$, 
and regard it as a representation of $\Spin_{2n}(\Comp)$ 
via the covering map $\pi : \Spin_{2n}(\Comp) \to \SOrth_{2n}(\Comp)$.
Let $e_1, \dots, e_{2n}$ be the standard basis of $\Comp^{2n}$.
If $1 \le r \le n-1$, then the exterior powers $\twedge^r \Comp^{2n}$ gives 
the irreducible highest weight representation of $\SOrth_{2n}(\Comp)$ or $\Spin_{2n}(\Comp)$ 
with highest weight $\ep_1 + \dots + \ep_r$ 
and highest weight vector $e_1 \wedge \cdots \wedge e_r$:
$$
\twedge^r \Comp^{2n} \cong V(\varpi_r)
\quad(1 \le r \le n-2),
\qquad
\twedge^{n-1} \Comp^{2n} \cong V(\varpi_{n-1} + \varpi_n).
$$
On the other hand, $\twedge^n \Comp^{2n}$ is not irreducible and
decomposes into two irreducible components:
$$
\twedge^n \Comp^{2n}
 \cong 
V(2 \varpi_n) \oplus V(2 \varpi_{n-1}),
$$
and the highest weight vector of $V(2 \varpi_n)$ (resp. $V(2 \varpi_{n-1})$) is 
given by $e_1 \wedge \cdots \wedge e_{n-1} \wedge e_n$ 
(resp. $e_1 \wedge \cdots \wedge e_{n-1} \wedge e_{n+1}$).
For a sequence $I = (i_1, \dots, i_r)$ of integers $i_k \in [2n] = \{ 1, 2, \dots, 2n \}$, 
we put
$$
e_I = e_{i_1} \wedge \cdots \wedge e_{i_r}.
$$
If $\binom{[2n]}{r}$ denotes the set of strictly increasing sequences 
$I = (i_1, \dots, i_r)$ of length $r$ with $i_1, \dots, i_r \in [2n]$, 
then $\{ e_I : I \in \binom{[2n]}{r} \}$ forms a basis of $\twedge^r \Comp^{2n}$.

Let $\langle \cdot, \cdot \rangle$ be the standard bilinear form on $\Comp^{2n}$ 
given by $\langle v, w \rangle = \trans v w$.
We define a bilinear form on $\twedge^r \Comp^{2n}$ by
$$
\langle v_1 \wedge \cdots \wedge v_r, w_1 \wedge \cdots \wedge w_r \rangle
 =
\det \left( \langle v_i, w_j \rangle \right)_{1 \le i, j \le r}.
$$
Then $\{ e_I : I \in \binom{[2n]}{r} \}$ forms an orthonormal basis 
of $\twedge^r \Comp^{2n}$.
If we denote by $\Delta^I_J(M)$ the minor of a $2n \times 2n$ matrix $M$ 
with rows indexed by $I = (i_1, \dots, i_r)$ 
and columns indexed by $J = (j_1, \dots, j_r)$, 
then we have
$$
\langle e_I, M \cdot e_J \rangle
 =
\Delta^I_J(M)
 =
\det \left( m_{i_p,j_q} \right)_{1 \le p, q \le r}.
$$
Note that this bilinear form satisfies the conditions in Lemma~\ref{lem:bilinear}.

The half-spin representations $V(\varpi_{n-1})$ and $V(\varpi_n)$ 
are minuscule ones 
(see \cite[VIII, \S7, n${}^\circ$3]{Bou:75} for properties of minuscule representations).
Let $\varpi = \varpi_{n-1}$ or $\varpi_n$.
If $S = \{ w \in W : w \varpi = \varpi \}$ is the stabilizer of $\varpi$ 
and $\{ w_1, \dots, w_N \}$ is a complete set of coset representatives for $W/S$, 
then $\{ \dot{w_i} v_\varpi : 1 \le i \le N \}$ forms a basis of $V(\varpi)$.
Moreover, by using the explicit action of $\g$ on $V(\varpi)$ given in 
\cite{Gec:17}, 
we can see that this is an orthonormal basis with respect to the bilinear form 
given in Lemma~\ref{lem:bilinear}.

In what follows we use the notation
$$
U(\tv',s)
 = 
\exp \Theta(\Lambda ; \tv',s)
 = 
\exp \left(
 \sum_{i=1}^{n-1} t_{2i-1} \Lambda^{2i-1}
  +
 c_n s \trans ( \hat{S \Lambda} ) \cdot S
 \right)
\in \SOrth_{2n}(\Comp).
$$
The determination of the $\tau$-functions are reduced to 
the computation of the minors of $U(\tv',s)$.

\begin{lemma}
\label{lem:tau=Delta}
\begin{enumerate}
\item[(1)]
If $1 \le r \le n-2$, then we have
$$
\tau_r(\Lambda;\tv',s;e_J)
 =
\Delta^{1,2,\dots,r}_{j_1,j_2,\dots, j_r} (U(\tv',s))
$$
for any $J = (j_1, j_2, \dots, j_r)$.
\item[(2)]
We have
\begin{gather*}
\left( \tau_n(\Lambda;\tv',s;\dot{w}\cdot v_{\varpi_n}) \right)^2
 =
\Delta^{1,\dots,n-1,n}_{w(1),\dots,w(n-1),w(n)} (U(\tv',s)),
\\
\left( \tau_{n-1}(\Lambda;\tv',s;\dot{w}\cdot v_{\varpi_{n-1}}) \right)^2
 =
\Delta^{1,\dots,n-1,n+1}_{w(1),\dots,w(n-1),w(n+1)} (U(\tv',s))
\end{gather*}
for any $w \in W$.
\end{enumerate}
\end{lemma}

\begin{proof}
We put $\tilde{U}(\tv',s) = \exp \Theta(\Lambda;\tv',s) \in \Spin_{2n}(\Comp)$.
Then the covering map $\pi : \Spin_{2n}(\Comp) \to \SOrth_{2n}(\Comp)$ 
sends $\tilde{U}(\tv',s)$ to $U(\tv',s)$.

(1)
Since the representation $V(\varpi_r)$ factors through $\SOrth_{2n}(\Comp)$, we have
$$
\tau_r ( \Lambda; \tv',s ; e_J )
 =
\langle e_I, \tilde{U}(\tv',s)\cdot  e_J \rangle_{\varpi_r}
 =
\langle e_I, U(\tv',s) \cdot e_J \rangle_{\varpi_r}
 =
\Delta^I_J (U(\tv',s)),
$$
where $I = (1, 2, \dots,  r)$.

(2)
We denote by $c^{\Spin_{2n}}_\lambda$ and $c^{\SOrth_{2n}}_\lambda$ 
the matrix coefficients on the groups $\Spin_{2n}(\Comp)$ and $\SOrth_{2n}(\Comp)$ 
respectively.
Let $r = n$ or $n-1$.
Then, by using \eqref{eq:mat_coeff3}, we have
\begin{align*}
\left( \tau_r(\Lambda;\tv',s; \dot{w}\cdot v_{\varpi_r}) \right)^2
 &=
\left(c^{\Spin_{2n}}_{\varpi_r}( \tilde{U}(\tv',s) \dot{w} )\right)^2
 =
c^{\Spin_{2n}}_{2\varpi_r}( \tilde{U}(\tv',s) \dot{w} )
 =
c^{\SOrth_{2n}}_{2\varpi_r}( U(\tv',s) \dot{w} )
\\
 &=
\langle v_{2 \varpi_r}, U(\tv',s) \dot{w}\cdot v_{2 \varpi_r} \rangle_{2\varpi_r}.
\end{align*}
Since $v_{2 \varpi_n} = e_{(1,\dots, n-1, n)}$ 
and $v_{2 \varpi_{n-1}} = e_{(1, \dots, n-1, n+1)}$, 
we obtain the desired formula.
\end{proof}

The key step of the proof of Theorem~\ref{thm:tau} is to express 
certain minors of $U(\tv',s)$ in terms of $2$-reduced Schur function.
Let $w \in W$, which is regarded as a permutation of $\{ 1, \dots, 2n \}$, 
and $J = J(w) = (j_1, \dots, j_n)$ be the sequence 
obtained by rearranging $w(1), \dots, w(n)$ in increasing order.
We associate to $w \in W$ the partition $\lambda = \lambda(w) = (\lambda_1, \dots, \lambda_n)$ 
given by
\begin{equation}
\label{eq:lambda}
\lambda_k
 = 
\begin{cases}
 j_{n+1-k} - (n+1-k) &\text{if $j_{n+1-k} \le n$,} \\
 j_{n+1-k} - (n+1-k)-1 &\text{if $j_{n+1-k} \ge n+1$.}
\end{cases}
\end{equation}
And we define $p = p(w)$ by
\[
p(w)
 = 
\# \{ k : j_k \ge n+2 \}
 = 
\# \{ k : \lambda_k \ge k \},
\]
that is, we have
\[
j_1<\cdots < j_{n-p}< n+2\le j_{n-p+1}<\cdots <j_n.
\]
Then $j_{n-p} = n$ or $n+1$, and 
$j_{n-p} = n$ (resp. $n+1$) exactly when $p$ is even (resp. odd).
\begin{lemma}
\label{lem:DeltaU=S}
Let $w \in W$, $J = J(w)$ and $\lambda = \lambda(w)$ be as above.
We put $I = (1,2, \dots, n)$.
\begin{enumerate}
\item[(1)]
If $\lambda_1 \le n-2$, then we have
$$
\Delta^I_J (U(\tv',s))
 = 
2^{p-1} S_{(\lambda_1, \dots, \lambda_n)}(\tv')
 +
2^{p-1} S_{(\lambda_1+1, \dots, \lambda_p+1, \lambda_{p+2}, \dots, \lambda_n)}(\tv').
$$
\item[(2)]
If $\lambda_1 = n-1$, then
\begin{align*}
\Delta^I_J (U(\tv',s))
 &= 
2^{p-1} S_{(\lambda_1, \dots, \lambda_n)}(\tv')
 +
2^{p-1} S_{(\lambda_1+1, \dots, \lambda_p+1, \lambda_{p+2}, \dots, \lambda_n)}(\tv')
\\
 &\quad
 +
(-1)^n s 
\left(
 2^{p-1} S_{(\lambda_1, \dots, \lambda_p, \lambda_{p+2}-1, \dots, \lambda_n-1)}(\tv')
 +
 2^{p-1} S_{(\lambda_2, \dots, \lambda_n)}(\tv')
\right)
\\
 &\quad
 +
s^2
\left(
 2^{p-2} S_{(\lambda_2-1, \dots, \lambda_n-1)}(\tv')
 +
 2^{p-2} S_{(\lambda_2, \dots, \lambda_p, \lambda_{p+2}-1, \dots, \lambda_n-1)}(\tv')
\right).
\end{align*}
\end{enumerate}
\end{lemma}

\begin{proof}
Let $V(\tv')$ be the $2n \times 2n$ matrix given by
$$
V(\tv')
 = 
U(\tv',0)
 =
\sum_{m=0}^{2n-2} q_m(\tv') \Lambda^m.
$$

First we expand the minor $\Delta^I_J (U(\tv',s))$ 
in terms of minors of $V(\tv')$.
Here we note that $i$ appears in $J$ if and only if $2n+1-i$ does not 
(see \eqref{eq:Weyl}).
If $j_1 = 1$, then by expanding $\Delta^I_J (U(\tv',s))$ along the first row, we have
$$
\Delta^{1,\dots,n}_{j_1,\dots,j_n} (U(\tv',s))
 =
\Delta^{1,\dots, n}_{j_1, \dots, j_n} (V(\tv'))
 +
(-1)^n s \cdot \Delta^{\hat{1}, 2, \dots, n}_{j_1, \dots, \hat{j_{n-p}}, \dots, j_n} (V(\tv')),
$$
where the symbol $\hat{i}$ means that we remove $i$ from the sequence.
Since the first column of the determinant 
$\Delta^{\hat{1}, 2, \dots, n}_{j_1, \dots, \hat{j_{n-p}}, \dots, j_n} (V(\tv'))$ 
is a zero vector, we have
$$
\Delta^{1,\dots,n}_{j_1,\dots,j_n} (U(\tv',s))
 =
\Delta^{1,\dots, n}_{j_1, \dots, j_n} (V(\tv')).
$$
If $j_n = 2n$, then by expanding $\Delta^I_J (U(\tv',s))$ along the first row 
and the last column, we have
\begin{align*}
\Delta^{1, \dots, n}_{j_1, \dots, j_n} (U(\tv',s))
 &=
\Delta^{1, \dots, n}_{j_1, \dots, j_n} (V(\tv'))
 +
(-1)^n s \cdot \Delta^{\hat{1},2, \dots, n}_{j_1, \dots, \hat{j_{n-p}}, \dots, j_n} (V(\tv'))
 +
(-1)^n s \cdot \Delta^{1, \dots, n-1,\hat{n}}_{j_1, \dots, j_{n-1},\hat{j_n}} (V(\tv'))
\\
 &\quad
 +
s^2 \cdot \Delta^{\hat{1}, 2, \dots, n}_{j_1, \dots, j_{n-1},\hat{j_n}} (V(\tv'))
 +
s^2 \cdot \Delta^{\hat{1},2, \dots, n-1,\hat{n}}_{j_1, \dots, \hat{j_{n-p}}, \cdots, j_{n-1},\hat{j_n}} (V(\tv')).
\end{align*}

Next we compute the above minors of $V(\tv')$ and express them 
in terms of $2$-reduced Schur functions.
Since the proofs are similar, we explain the computation of 
$\Delta^{1, \dots, n}_{j_1, \dots, j_n} (V(\tv'))$.
In fact, we prove
\begin{equation}
\label{eq:Delta=S}
\Delta^{1, \dots, n}_{j_1, \dots, j_n} (V(\tv'))
 =
2^{p-1} S_{(\lambda_1, \dots, \lambda_n)}(\tv')
 +
2^{p-1} S_{(\lambda_1+1, \dots, \lambda_p+1, \lambda_{p+2}, \dots, \lambda_n)}(\tv').
\end{equation}

Consider the case $j_{n-p} = n$.
In this case, by multiplying the last column by $1/2$ and the last row by $2$, 
we see that
$$
\Delta^{1, 2, \dots, n}_{j_1,j_2, \dots, j_n} (V(\tv'))
 =
\det \begin{pmatrix}
 q_{\lambda_n} & \cdots & q_{\lambda_{p+2}+n-p-2} & q_{n-1}
  & 2 q_{\lambda_p+n-p} & \cdots & 2 q_{\lambda_2+n-2} & q_{\lambda_1+n-1} \\
               & \ddots &                         & \vdots 
  & \vdots              &        &                     & \vdots \\
               &        & q_{\lambda_{p+2}}       & q_{\lambda_{p+1}+1} 
  & \vdots              &        &                     & \vdots \\
               &        &                         & q_{\lambda_{p+1}}
  & 2 q_{\lambda_p+1}   &        &                     & \vdots \\
               &        &                         & q_{\lambda_{p+1}-1}
  & 2 q_{\lambda_p}     & \ddots &                     & \vdots \\
               &        &                         & \vdots
  & \vdots              & \ddots & \ddots & \vdots \\
               &        &                         & q_1
  & 2 q_{\lambda_p-p+2} &        & 2 q_{\lambda_2}     &  q_{\lambda_1+1} \\[5pt]
 0             & \cdots & 0                       & 2 
  & 2 q_{\lambda_p-p+1} & \cdots & 2 q_{\lambda_2-1}   & q_{\lambda_1}
\end{pmatrix}.
$$
By splitting the last row as the sum of two vectors
$$
\begin{matrix}
a = &
 ( & 0 & \cdots & 0 & 1 & 2 q_{\lambda_p-p+1} & \cdots & 2 q_{\lambda_2-1} & q_{\lambda_1}
 & ),
 &\text{and} \\
b = &
 ( & 0 & \cdots & 0 & 1 & 0                   & \cdots & 0                 & 0
 & ),
\end{matrix}
$$
we obtain
$$
\Delta^{1, 2, \dots, n}_{j_1, \dots, j_n} (V(\tv'))
 =
\det A + \det B,
$$
where $A$ and $B$ are the matrices obtained by replacing the last row 
with $a$ and $b$ respectively.
Comparison with the definition (\ref{eq:Schur}) gives $\det A = 2^{p-1} S_\lambda(\tv')$.
And, by expanding $\det B$ along the last row, 
we see 
$\det B
 = 2^{p-1} S_{(\lambda_1+1, \dots, \lambda_p+1, \lambda_{p+2}, \dots, \lambda_n)}(\tv')$.

If $j_{n-p} = n+1$, then we have
$$
\Delta^{1, 2, \dots, n}_{j_1,j_2, \dots, j_n} (V(\tv'))
 =
\det \begin{pmatrix}
 q_{\lambda_n} & \cdots & q_{\lambda_{p+2}+n-p-2} & q_{n-1}
  & 2 q_{\lambda_p+n-p} & \cdots & 2 q_{\lambda_2+n-2} & q_{\lambda_1+n-1} \\
               & \ddots &                         & \vdots 
  & \vdots              &        &                     & \vdots \\
               &        & q_{\lambda_{p+2}}       & q_{\lambda_{p+1}+1} 
  & \vdots              &        &                     & \vdots \\
               &        &                         & q_{\lambda_{p+1}}
  & 2 q_{\lambda_p+1}   &        &                     & \vdots \\
               &        &                         & q_{\lambda_{p+1}-1}
  & 2 q_{\lambda_p}     & \ddots &                     & \vdots \\
               &        &                         & \vdots
  & \vdots              & \ddots & \ddots & \vdots \\
               &        &                         & q_1
  & 2 q_{\lambda_p-p+2} &        & 2 q_{\lambda_2}     &  q_{\lambda_1+1} \\[5pt]
 0             & \cdots & 0                       & 0 
  & 2 q_{\lambda_p-p+1} & \cdots & 2 q_{\lambda_2-1}   & q_{\lambda_1}
\end{pmatrix}.
$$
By splitting the last row as the difference of two vectors
$$
\begin{matrix}
a = &
 ( & 0 & \cdots & 0 & 1 & 2 q_{\lambda_p-p+1} & \cdots & 2 q_{\lambda_2-1} & q_{\lambda_1}
 & ),
 &\text{and} \\
b = &
 ( & 0 & \cdots & 0 & 1 & 0                   & \cdots & 0                 & 0
 & ),
\end{matrix}
$$
we obtain (\ref{eq:Delta=S}).

Similarly we can show
\begin{align*}
\Delta^{\hat{1},2, \dots, n}_{j_1, \dots, \hat{i_{n-p}}, \dots, j_n} (V(\tv'))
 &=
2^{p-1} S_{(\lambda_1, \dots, \lambda_p, \lambda_{p+2}-1, \dots, \lambda_n-1)}(\tv'),
\\
\Delta^{1, \dots, n-1,\hat{n}}_{j_1, \dots, j_{n-1},\hat{j_n}} (V(\tv'))
 &=
2^{p-1} S_{(\lambda_2, \dots, \lambda_n)}(\tv'),
\\
\Delta^{\hat{1},2, \dots, n}_{j_1, \dots, j_{n-1,\hat{j_n}}} (V(\tv'))
 &=
2^{p-2} S_{(\lambda_2-1, \dots, \lambda_n-1)}(\tv')
-
2^{p-2} S_{(\lambda_2, \dots, \lambda_p, \lambda_{p+2}-1, \dots, \lambda_n-1)}(\tv'),
\\
\Delta^{\hat{1}, 2, \dots, n-1,\hat{n}}_{j_1, \dots, \hat{j_{n-p}}, \dots, j_{n-1},\hat{j_n}} (V(\tv'))
 &=
2^{p-1} S_{(\lambda_2, \dots, \lambda_p, \lambda_{p+2}-1, \dots, \lambda_n-1)}(\tv').
\end{align*}
Combining these formulas for the minors of $V(\tv')$, we complete the proof.
\end{proof}

The following lemma enables us to relate the partitions appearing 
in Lemma~\ref{lem:DeltaU=S} with the strict partition $\alpha(w)$ 
given by (\ref{eq:alpha}).
Given a partition $\mu$, we define
$$
l = \# \{ i : \mu_i \ge i \},
\qquad
\beta_i = \mu_i - i,
\quad
\gamma_i = \trans\mu_i - i
\quad(1 \le i \le l),
$$
where $\trans\mu$ is the conjugate partition of $\mu$, 
and write $\mu = (\beta_1, \dots, \beta_l | \gamma_1, \dots, \gamma_l)$.
This representation is called the Frobenius notation of $\mu$.

\begin{lemma}
\label{lem:lambda-alpha}
Let $w \in W$ and $\lambda = \lambda(w)$ the partition defined by (\ref{eq:lambda}).
Then we have
$$
\trans\lambda_k = \lambda_k + 1 \quad(1 \le k \le p),
\qquad
\lambda_{p+1} = p,
\qquad
\trans\lambda_k = \lambda_{k+1} \quad(p+1 \le k \le n-1).
$$
If $\alpha = \alpha(w)$ is the strict partition given by \eqref{eq:alpha}, 
then we have $\alpha_k = \lambda_k -k+1$ for $1 \le k \le p$ and 
\begin{align*}
(\lambda_1, \dots, \lambda_n)
 &=
(\alpha_1 - 1, \dots, \alpha_p - 1 | \alpha_1, \dots, \alpha_p),
\\
(\lambda_1+1, \dots, \lambda_p+1, \lambda_{p+2}, \dots, \lambda_n)
& =
(\alpha_1, \dots, \alpha_p | \alpha_1 - 1, \dots, \alpha_p - 1),
\\
(\lambda_1, \dots, \lambda_p, \lambda_{p+2}-1, \dots, \lambda_n-1)
 &=
(\alpha_1 - 1, \dots, \alpha_p - 1 | \alpha_2, \dots, \alpha_p,0),
\\
(\lambda_2, \dots, \lambda_n)
& =
(\alpha_2, \dots, \alpha_p, 0 | \alpha_1 - 1, \dots, \alpha_p - 1),
\\
(\lambda_2-1, \dots, \lambda_n-1)
& =
(\alpha_2 - 1, \dots, \alpha_p - 1 | \alpha_2, \dots, \alpha_p),
\\
(\lambda_2, \dots, \lambda_p, \lambda_{p+2}-1, \dots, \lambda_n-1)
& =
(\alpha_2, \dots, \alpha_p | \alpha_2 - 1, \dots, \alpha_p - 1).
\end{align*}
\end{lemma}

\begin{proof}
Since $j_{n-p} = n$ or $n+1$, we can see that $\lambda_{p+1} = p$.

We prove $\trans\lambda_k = \lambda_k+1$ for $1 \le k \le p$.
The largest $p$ elements of $J$ are
$$
j_{n+1-k}
 =
\lambda_k+n-k+2
\qquad(1 \le k \le p).
$$
Since $i \in J$ if and only if $2n+1-i \not\in J$, we see that 
the smallest $p$ elements of $J^c = \{ 1, \dots, 2n \} \setminus J$ are
$$
(2n+1) - (\lambda_k+n-k+2) = n+k-1 - \lambda_k
\qquad(1 \le k \le p).
$$
On the other hand, it follows from \cite[I.(1.7)]{Mac:95} that 
$$
\{ \lambda_i -i : 1 \le i \le p \}
 \sqcup
\{ -1+j-\trans\lambda_j : p+1 \le j \le n-1 \}
 =
\{ 0, 1, \dots, n-2 \},
$$
and that the smallest $p$ elements of $J^c$ are
$$
n+k - \trans\lambda_k
\qquad(1 \le k \le p).
$$
Hence we obtain $\trans\lambda_k = \lambda_k+1$ for $1 \le k \le p$.

Similarly, by considering the largest $(n-1-p)$ elements of $J^c$, 
we can prove $\trans\lambda_k = \lambda_{k+1}$ for $p+1 \le k \le n-1$.

The latter half of the lemma follows from the definition 
of the Frobenius notation.
\end{proof}

The last ingredient of the proof of Theorem~\ref{thm:tau} 
is the following relations between $2$-reduced Schur functions and Schur's $Q$-functions.

\begin{lemma}
\label{lem:relationSQ}
\begin{enumerate}
\item[(1)]
(see \cite[III.8 Example~10 (a)]{Mac:95})
For any partition $\lambda$, we have
$$
S_\lambda(\tv')
 =
S_{\trans\lambda}(\tv').
$$
\item[(2)]
(\cite[(15)]{You:89}, \cite[Theorem~2]{JP:91})
If $\lambda = (\alpha_1, \dots, \alpha_p | \alpha_1 - 1, \dots, \alpha_p - 1)$, then we have
$$
S_{(\alpha_1, \dots, \alpha_p|\alpha_1-1, \dots, \alpha_p-1)}(\tv')
 =
2^{-p}
\left(Q_{(\alpha_1, \dots, \alpha_p)}(\tv')\right)^2.
$$
\item[(3)]
(\cite[Theorem~1.3]{You:92})
If $\lambda = (\alpha_1-1, \dots, \alpha_p-1 | \alpha_2, \dots, \alpha_p, 0)$, then we have
$$
S_{(\alpha_1-1, \dots, \alpha_p-1 | \alpha_2, \dots, \alpha_p, 0)}(\tv')
 =
2 \cdot 2^{-p}
Q_{(\alpha_1, \dots, \alpha_p)}(\tv')
Q_{(\alpha_2, \dots, \alpha_p)}(\tv').
$$
\end{enumerate}
\end{lemma}

Now we are in position to finish the proof of Theorem~\ref{thm:tau}.

\begin{proof}[Proof of Theorem~\ref{thm:tau}]
First we prove that $\tau_n(\Lambda;\tv;\dot{w}\cdot v_{\varpi_n})
 = c_{n,w} \hat{Q}_{\alpha(w)}(\tv',s)$ (for some constant $c_{n,w}$) 
(Equation (\ref{eq:tau=Q1})).
By Lemma~\ref{lem:tau=Delta} (2), we have
$$
\left(\tau_n(\Lambda;\tv;\dot{w}\cdot v_{\varpi_n})\right)^2
 =
\pm \Delta^I_J (U(\tv',s)),
$$
where $I = (1, 2, \dots, n)$ and $J$ is the rearrangement of $\{ w(1), w(2), \dots, w(n) \}$ 
in increasing order.

We consider the case where $n-1$ is a part of $\alpha(w)$, 
i.e. $\lambda_1 = \alpha_1 = n-1$.
Then, combining Lemmas~\ref{lem:DeltaU=S}, \ref{lem:lambda-alpha} 
and \ref{lem:relationSQ}, we obtain
\begin{align*}
\Delta^I_J (U(\tv',s))
 &=
2^p S_{(\alpha_1-1,\dots,\alpha_p-1|\alpha_1,\dots,\alpha_p)}(\tv')
\\
 &\quad
 +
2^p \cdot (-1)^n s \cdot 
S_{(\alpha_1-1,\dots,\alpha_p-1|\alpha_2,\dots,\alpha_p,0)}(\tv')
\\
 &\quad
 +
2^{p-1} \cdot s^2 \cdot
S_{(\alpha_2-1,\dots,\alpha_p-1|\alpha_2,\dots,\alpha_p)}(\tv')
\\
 &=
Q_{(\alpha_1, \dots, \alpha_p)}(\tv')^2
 +
2 (-1)^n s Q_{(\alpha_1, \dots, \alpha_p)}(\tv') Q_{(\alpha_2,\dots, \alpha_p)}(\tv')
 +
s^2 \left(Q_{(\alpha_2, \dots, \alpha_p)}(\tv')\right)^2
\\
 &=
\left(\hat{Q}_\alpha(\tv',s)\right)^2.
\end{align*}
Similarly we can show the case where $n-1$ is not a part of $\alpha(w)$.

Equation (\ref{eq:tau=Q2}) can be derived from (\ref{eq:tau=Q1}) 
by using the Dynkin diagram automorphism of $\sorth_{2n}(\Comp)$.
Let $\sigma \in \Orth_{2n}(\Comp)$ be the matrix given by
$$
\sigma = 
\begin{pmatrix}
 I_{n-1} &   &   &         \\
         & 0 & 1 &         \\
         & 1 & 0 &         \\
         &   &   & I_{n-1}
\end{pmatrix}.
$$
Then $\sigma^2 = 1$ and the conjugation by $\sigma$ induces the Dynkin diagram automorphism 
switching the simple roots $\alpha_{n-1} = \ep_{n-1} - \ep_n$ 
and $\alpha_n = \ep_{n-1} + \ep_n$.
And the induced automorphism on $W \cong \WD_n$ is given by 
$\WD_n \ni w \mapsto w^\dagger = (n,n+1) w (n,n+1) \in \WD_n$.
Since we have
$$
\sigma \cdot v_{2 \varpi_n}
 =
\sigma \cdot e_1 \wedge \cdots \wedge e_{n-1} \wedge e_n
 =
e_1 \wedge \cdots \wedge e_{n-1} \wedge e_{n+1}
 =
v_{2 \varpi_{n-1}},
$$
it follows from Lemma~\ref{lem:tau=Delta} that
\begin{align*}
\left(\tau_{n-1} (\Lambda;\tv;\dot{w}\cdot v_{2 \varpi_{n-1}})\right)^2
 &=
\langle v_{2 \varpi_{n-1}}, U(\tv';s) \dot{w}\cdot v_{2 \varpi_{n-1}} \rangle
 =
\langle \sigma\cdot v_{2 \varpi_n}, U(\tv',s) \dot{w} \sigma \cdot v_{2 \varpi_n} \rangle
\\
 &=
\langle 
 e_I,
 \sigma^{-1} U(\tv',s) \sigma \cdot \sigma^{-1} \dot{w} \sigma \cdot e_I
\rangle.
\end{align*}
Here we note that $\sigma^{-1} U(\tv',s) \sigma$ is the matrix obtained 
from $U(\tv',s)$ by swapping the $n$th and $(n+1)$st rows and 
the $n$th and $(n+1)$st columns.
Then we have
$$
\sigma^{-1} U(\tv',s) \sigma = U(\tv',-s),
$$
and (\ref{eq:tau=Q2}) is obtained from (\ref{eq:tau=Q1}).
This completes the proof of Theorem~\ref{thm:tau}.
\end{proof}

\begin{remark}
The argument in the proof of Theorem~\ref{thm:tau} shows that
the relations for the matrix coefficients (see Lemma~\ref{lem:mat_coeff} (3))
\begin{gather*}
c_{2\varpi_n}( U(\tv',s) \dot{w} )
 =
\langle
 e_1 \wedge \cdots \wedge e_{n-1} \wedge e_n,
 U(\tv',s) \dot{w} \cdot e_1 \wedge \cdot \wedge e_{n-1} \wedge e_n
\rangle
 =
\left( c_{\varpi_n}( U(\tv',s) \dot{w} ) \right)^2,
\\
c_{2\varpi_{n-1}}( U(\tv',s) \dot{w} )
 =
\langle
 e_1 \wedge \cdots \wedge e_{n-1} \wedge e_{n+1},
 U(\tv',s) \dot{w} \cdot e_1 \wedge \cdot \wedge e_{n-1} \wedge e_{n+1}
\rangle
 =
\left( c_{\varpi_{n-1}}( U(\tv',s) \dot{w} ) \right)^2
\end{gather*}
are translated into bilinear expansion formulas of $2$-reduced Schur functions 
in Schur's $Q$-functions, which are the identities in Lemma~\ref{lem:relationSQ} (2) and (3).
Similarly, since $\varpi_{n-1} + \varpi_n = \ep_1 + \dots + \ep_{n-1}$, 
it follows from Lemma~\ref{lem:mat_coeff} (3) that
$$
c_{\varpi_{n-1} + \varpi_n}( U(\tv',s) \dot{w} )
 =
\langle
 e_1 \wedge \cdots \wedge e_{n-1},
 U(\tv',s) \dot{w} \cdot  e_1 \wedge \cdots \wedge e_{n-1} 
\rangle
 =
c_{\varpi_{n-1}}( U(\tv',s) \dot{w} )
\cdot
c_{\varpi_{n}}( U(\tv',s) \dot{w} ).
$$
By using Theorem~\ref{thm:tau} and a similar argument to the proof of Lemma~\ref{lem:DeltaU=S}, 
we can derive some bilinear expansion formulas.
For example, if $n+2 \le w(n) \le 2n-1$, equating the coefficients of $s$ gives us
\begin{multline*}
2^{p-1} S_{(\alpha_1, \dots, \hat{\alpha_i}, \dots, \alpha_p | \alpha_2-1, \dots, \alpha_p-1)}(\tv')
\\
 =
 -
Q_{(\alpha_1, \dots, \alpha_p)}(\tv')
Q_{(\alpha_2, \dots, \hat{\alpha_i}, \dots, \alpha_p)}(\tv')
 +
Q_{(\alpha_1, \dots, \hat{\alpha_i}, \dots, \alpha_p)}(\tv')
Q_{(\alpha_2, \dots, \alpha_p)}(\tv'),
\end{multline*}
which is a special case of \cite[Corollary~2]{P:98}.
See \cite{JP:91, P:98, You:92, HO:21a} for bilinear expansion formulas 
of $2$-reduced Schur functions in Schur's $Q$-functions.
\end{remark}

We also have the following remark on Theorem \ref{thm:tau}.

\begin{remark}
Consider the fixed-point subgroup of the diagram automorphism $\dagger : W \to W$:
\[
W'
 =
\{ w \in W : w = w^\dagger \},
\]
which is isomorphic to the Weyl group of type $B_{n-1}$.
Then, for $w \in W'$ and $s=0$, we have
\[
\tau_n(\Lambda;\tv',0;\dot w\cdot v_{\varpi_n})
=
\tau_{n-1}(\Lambda;\tv',0;\dot w\cdot v_{\varpi_{n-1}})
=
Q_{\alpha(w)}(\tv')
\]
up to constant multiples.
This is the polynomial $\tau$-function of the f-KT hierarchy on $\sorth_{2n-1}(\Comp)$ 
of type $B_{n-1}$ corresponding to the spin node \cite{Xie:21}.
Also note that both $\tau_{n-1}(\Lambda;\tv',0;\dot{w}\cdot v_{\varpi_{n-1}})$ 
and $\tau_n(\Lambda;\tv',0;\dot{w}\cdot v_{\varpi_{n}})$ satisfy the BKP hierarchy, 
and we expect that all the polynomial $\tau$-functions of the BKP hierarchy 
are given by these $\tau$-functions with $w \in W'$ \cite{You:89, KL:19, HO:21, Le:22}.
\end{remark}

\subsection{%
Explicit formulas for $\tau_r$ with $1 \le r \le n-2$
}

In this final subsection, we give explicit formulas for 
the $\tau_r$-functions with $1 \le r \le n-2$ in terms of 
$2$-reduced Schur functions.

\begin{proposition}
\label{prop:tau}
Let $J = (j_1, \dots, j_r) \in \binom{[2n]}{r}$ 
and regard it as a subset of $[2n]=\{ 1, \dots, 2n \}$.
We define a sequence $\lambda = (\lambda_1, \dots, \lambda_r)$ by
$$
\lambda_k
 =
\begin{cases}
 j_{r-k+1} - (r-k+1) - 1 &\text{if $j_{r-k+1} \ge n+1$,} \\
 j_{r-k+1} - (r-k+1)     &\text{if $j_{r-k+1} \le n$}
\end{cases}
$$
and $p = \# \{ k : j_k \ge n+2 \}$.
Then the $\tau_r$-functions for $1\le r\le n-2$ are given as follows 
(up to nonzero constant multiples):
\begin{enumerate}
\item[(1)]
If $J \cap \{ n, n+1, 2n \} = \emptyset$, then we have
$$
\tau_r(\Lambda;\tv;e_J)
 =
2^p S_\lambda(\tv').
$$
\item[(2)]
If $J \cap \{ n, n+1, 2n \} = \{ 2n \}$, then we have
$$
\tau_r(\Lambda;\tv;e_J)
 =
2^p S_\lambda(\tv')
 + 
(-1)^{n+r} 2^{p-1} s^2 S_{(\lambda_2-1, \dots, \lambda_r-1)}(\tv').
$$
\item[(3)]
If $J \cap \{ n, n+1, 2n \} = \{ n+1 \}$, then we have
$$
\tau_r(\Lambda;\tv;e_J)
 =
2^p S_\lambda(\tv')
 + 
(-1)^{r-p+1} 2^p s
 S_{(\lambda_1, \dots, \lambda_p, \lambda_{p+2}-1, \dots, \lambda_r-1)}(\tv').
$$
\item[(4)]
If $J \cap \{ n, n+1, 2n \} = \{ n \}$, then we have
$$
\tau_r(\Lambda;\tv;e_J)
 =
2^p S_\lambda(\tv')
 +
(-1)^{r-p} 2^p s
 S_{(\lambda_1, \dots, \lambda_p, \lambda_{p+2}-1, \dots, \lambda_r-1)}(\tv').
$$
\item[(5)]
If $J \cap \{ n, n+1, 2n \} = \{ n+1, 2n \}$, then we have
$$
\tau_r(\Lambda;\tv;e_J)
 =
2^p S_\lambda(\tv')
 +
(-1)^{r-p+1} 2^p s
 S_{(\lambda_1, \dots, \lambda_p, \lambda_{p+2}-1, \dots, \lambda_r-1)}(\tv')
 +
(-1)^{n+r} 2^{p-1} s^2
 S_{(\lambda_2-1, \dots, \lambda_r-1)}(\tv').
$$
\item[(6)]
If $J \cap \{ n, n+1, 2n \} = \{ n, 2n \}$, then we have
$$
\tau_r(\Lambda;\tv;e_J)
 =
2^p S_\lambda(\tv')
 +
(-2)^{r-p} 2^p s
 S_{(\lambda_1, \dots, \lambda_p, \lambda_{p+2}-1, \dots, \lambda_r-1)}(\tv')
 +
(-1)^{n+r} 2^{p-1} s^2
 S_{(\lambda_2-1, \dots, \lambda_r-1)}(\tv').
$$
\item[(7)]
If $J \cap \{ n, n+1, 2n \} = \{ n, n+1 \}$, then we have
$$
\tau_r(\Lambda;\tv;e_J)
 =
(-1)^{r-p+1} 2^{p+1} s
 S_{(\lambda_1, \dots, \lambda_p, \lambda_{p+2}-1, \dots, \lambda_r-1)}(\tv').
$$
\item[(8)]
If $J \cap \{ n, n+1, 2n \} = \{ n, n+1, 2n \}$, then we have
$$
\tau_r(\Lambda;\tv;e_J)
 =
(-1)^{r-p+1} 2^{p+1} s 
 S_{(\lambda_1, \dots, \lambda_p, \lambda_{p+2}-1, \dots, \lambda_r-1)}(\tv').
$$
\end{enumerate}
\end{proposition}

\begin{proof}
We can prove this proposition in a similar manner to the proof of Lemma~\ref{lem:DeltaU=S}.
So we omit it.
\end{proof}

\raggedright



\end{document}